\documentclass[aps,pre,notitlepage]{revtex4-1}
\usepackage{amsmath,amssymb,bm}
\usepackage{graphicx}
\usepackage{subfigure}
\begin{document}
\title{
Absorbing phase transition in the coupled dynamics of node and link states in random networks
}
\author{Meghdad Saeedian,Maxi San Miguel,Raul Toral}
\affiliation{IFISC Instituto de Fisica Interdisciplinar y Sistemas Complejos (CSIC-UIB), Campus Universitat Illes Balears, E-07122 Palma de Mallorca, Spain}
\begin{abstract}
We present a stochastic dynamics model of coupled evolution for the binary states of nodes and links in a complex network. In the context of opinion formation node states represent two possible opinions and link states a positive or negative relation. Dynamics proceeds via node and link state update towards pairwise satisfactory relations in which nodes in the same state are connected by positive links or nodes in different states are connected by negative links. By a mean-field rate equations analysis and Monte Carlo simulations in random networks we find an absorbing phase transition from a dynamically active phase to an absorbing phase. The transition occurs for a critical value of the relative time scale for node and link state updates. In the absorbing phase the order parameter, measuring global order, approaches exponentially the final frozen configuration. Finite size effects are such that in the absorbing phase the final configuration is reached in a characteristic time that scales logarithmically with system size, while in the active phase, finite-size fluctuation take the system to a frozen configuration in a characteristic time that grows exponentially with system size. There is also a finite-size topological transition associated with group splitting in the network of these final frozen configurations.
\end{abstract}
\maketitle

\section{Introduction}
Complex networks are the skeletons of complex systems. The dynamical properties of these systems are normally studied by considering the changes in the states of the nodes as a consequence of the interaction with their neighbors in the network. More recently, but with earlier analysis in social sciences \cite{heider1946attitudes}, there has been a focus on dynamical problems associated with states of the links\cite{radicchi2007social,szell2010multirelational,marvel2011continuous, traag2009community,evans2009line,ahn2010link,nepusz2012controlling, antal2005dynamics,marvel2009energy,antal2006social,leskovec2010signed,fernandez2012dynamics,carro2014fragmentation,shi2016evolution}. However, little attention \cite{carro2016coupled,saeedian2017epidemic,singh2014extreme}has been paid to problems in which both node and link states are taken into account in a coupled dynamics: The state of the nodes and the nature of their interactions dynamically update each other. Hence the states of the nodes condition the states of the links as much as the states of the links condition the states of the nodes. This coevolution of states and interactions mediated by links is at the heart of the complexity \cite{holovatch2017complex} of social systems, where positive or negative interactions are associated with concepts such as friendship, trust, etc. In this paper we consider this general situation in a model of opinion formation with a binary choice for both the state of the node and the state of the link.

The importance of the state of the link in social networks has been emphasized in the context of triangular relations with friendly or unfriendly links. Historically, a first theory about psychological three-person relations is the structural balance theory introduced by Heider\cite{heider1946attitudes} in 1946. This theory exemplifies the principles that ``\textit{the friend (enemy) of my friend (enemy) is my friend}'' and ``\textit{the friend (enemy) of my enemy (friend) is my enemy}''. Later, it was translated to the language of graphs\cite{cartwright1956structural} and has been a scientific challenge in the study of social systems \cite{radicchi2007social,szell2010multirelational,marvel2011continuous,antal2005dynamics,marvel2009energy,antal2006social,saeedian2017epidemic,leskovec2010signed}. In particular Leskovec \textit{et al}\cite{leskovec2010signed} concluded that Heider balance theory cannot explain the observed triangular patterns in a large data set, and proposed the status theory as an alternative explanation of triangle formation in networks. Both theories predict different signs (positive or negative, meaning, respectively, friendly or unfriendly) for a given link in some triangles. In these studies only the state of the link is considered and the question addressed is about the structural balance of triangles. A step beyond these studies are those, either in opinion formation (friendly or unfriendly link) \cite{shi2016evolution} or epidemics (active or inert link)\cite{kermack1932contributions,kermack1933contributions}, in which both nodes and links have a state, but the state of the link is either fixed or, else, determined by the state of the connecting nodes, so that, still, there is no coupled dynamics of node and link states.

The dynamical interplay between the states of the nodes and the state of the links has been considered in the context of an Ising model \cite{singh2014extreme} and a susceptible infected model\cite{saeedian2017epidemic}. In both cases the dynamics to approach structural balance is described by energy minimization of an appropriate Hamiltonian in a fully connected network. A different approach is that of Carro \textit{et al} \cite{carro2016coupled} in the context of language competition (node state: language preference, link state: language use) where a genuine non-equilibrium dynamics with no Hamiltonian minimization is implemented in a complex network. This leads to a wide range of asymptotic states, including long-lived dynamically states. Here we also consider a non-equilibrium dynamics in the context of opinion formation and we focus on binary or pairwise relations: Nodes can be in either of two states or opinions and links can be positive (friendly) or negative (unfriendly). Satisfactory pairwise relations are those of friendly links connecting nodes in the same state or unfriendly links connecting nodes in different states. The basic assumption is that unsatisfying pairs evolve to satisfying ones either by updating the state of the link or by updating the state of one of the nodes (see Fig.~\ref{update_rule}).

As a main result of our study, we find an absorbing transition from a dynamically active state to an absorbing frozen configuration in which all pairwise relations are satisfactory. The absorbing phase occurs beyond a critical value $p_c$ of the parameter that measures the ratio of time scales for links and nodes updates. This implies that, despite a dynamical rule of local convergence, a global convergence to the absorbing state only occurs when the states of the links evolve fast enough in comparison with the evolution of the states of the nodes. The two phases separated by $p_c$ can be also characterized dynamically: From general random initial conditions and for $p>p_c$ the evolution of an order parameter that measures global ordering indicates an exponential approach to the absorbing state, while for $p<p_c$ the order parameter falls into a plateau value characterizing partial ordering of the system. Finite size effects lead to very different time scales: In the absorbing phase the frozen configuration is reached in a characteristic time that scales logarithmically with system size, while in the active phase finite-size fluctuations bring the system to a frozen configuration in a characteristic time that scales exponentially with system size. In addition, we find a transition associated with the topology of frozen configuration of the absorbing phase and the frozen configuration reached by finite-size fluctuations in the active phase. This is a finite-size transition that disappears in the thermodynamic limit.

This paper is organized as follows. After introducing our dynamical model, we discuss a mean-field rate equation approach, suitable for homogeneous random networks \cite{barrat2008dynamical}, that is based on the assumption that each node has exactly $\mu$ neighbors. These equations, that neglect fluctuations and are only valid in the thermodynamic limit, predict a continuous transition between absorbing and active phases and allow the calculation of $p_c$ as a function of the mean degree of the network $\mu$. Next we report Monte Carlo simulations of the model on Erd\"os-R\'enyi networks that confirm the rate equation predictions, and describe the dynamical properties of the active and frozen phase, including finite-size effects. We finally summarize our results.

\section{Results}

\section{Joint-evolution of node and links property in the imitating process}
\begin{figure}[h]\centering
\includegraphics[width=0.6\linewidth]{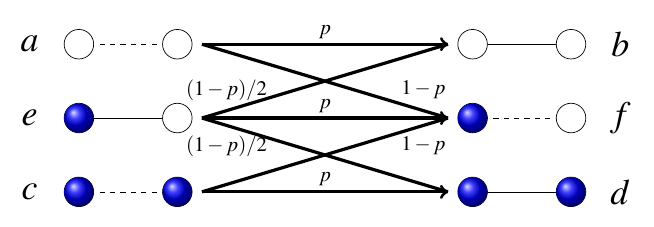}
\caption{We present in this figure all six possible configurations of pairs, where pairs $a$, $c$ and $e$ are unsatisfying while pairs $b$, $d$ and $f$ are satisfying. Moreover, we also depict the dynamical rules that turn unsatisfying pairs into satisfying one through node or link updates. \textit{p} is the probability of link update and $1-p$, the complementary probability, is the probability of node update.}
\label{update_rule}
\end{figure}

Let us consider a network defined as a set of nodes and links. The nodes represent individuals and the links, understood as undirected connections, indicate a relation between the nodes. Each node holds a binary state variable whose value represents one of two possible opinions. In the figures those two possible values are indicated by a dark (blue) or white color. The links between nodes represent one of the two possible types of relationship: friendly (attraction) and unfriendly (repulsion). In the figures they are indicated, respectively, by a continuous or a dashed line. According to the aforementioned interpretation, we consider that friendly links between pairs of nodes holding the same opinion or unfriendly links between pairs of nodes holding different opinions are {\bfseries satisfying} links (or form satisfying pairs), while friendly links between nodes holding different opinions or unfriendly links between nodes holding the same opinion are {\bfseries unsatisfying}. All possible situations are displayed in Fig.~\ref{update_rule}: pairs \textit{a}, \textit{c} and \textit{e} are unsatisfying, while pairs \textit{b}, \textit{d} and \textit{f} are satisfying.

Our basic assumption is that people in a community act in order to maximize their level of satisfaction, a desirable option from the psychological viewpoint. Hence, individuals belonging to an unsatisfying pair connection take action to turn it into satisfying. To convert unsatisfying pairs into satisfying ones, individuals can either change their opinions or alter the link state. Consider, for instance, the unsatisfying \textit{a} pair in Fig.~\ref{update_rule}. It can become satisfying if one of the individuals changes its opinion state from white to blue or, alternatively, if they decide to change their link state from unfriendly to friendly. We consider that these two options happen with probability $1-p$ and $p$, respectively. A similar scenario holds for the unsatisfying link \textit{c}. In the case \textit{e} we need to assign a probability $(1-p)/2$ to the node holding the blue opinion changing it to white and a probability $(1-p)/2$ to the node holding the white opinion changing it to blue. In the pairs \textit{a, c}, it does not matter which one node changes opinion, as it always leads to an \textit{f} pair.

In a Monte Carlo implementation of these dynamical rules, a link is chosen at random from all existing links. If the corresponding pair is satisfying, nothing happens; otherwise, if the pair is unsatisfying, it is converted into satisfying by applying the rules, with their respective probabilities, displayed in Fig.~\ref{update_rule}. A Monte Carlo step, as usual, is defined as a number of consecutive link selections equal to the total number of links existing in the system. Most of our results concern an uncorrelated Erd\"os-R\'enyi network with $N$ nodes and average degree (number of links per node) $\mu$ such that the total number of links is $L=\frac12 \mu N$. Defining $L_i$, \textit{$i\in$\{a, b, c, d, e, f\}} as the number of links of type $i$, the associated densities are $\rho_i=L_i/L$, satisfying the obvious normalization condition $\rho_{a}+\rho_{b}+\rho_{c}+\rho_{d}+\rho_{e}+\rho_{f}=1$.

When all pairs are satisfying, no further evolution is possible and the system is dynamically frozen in an absorbing state. Note that an update that converts a pair from unsatisfying to satisfying by changing an individual node state might change the status of another pair, to which the node involved in the update also belongs to, from satisfying to unsatisfying. Therefore, we ask the question of under which conditions the dynamical rules defined above lead to an absorbing, all pairs being satisfying, global state.

In the frozen state, the densities $\rho_{a},\,\rho_{c},\,\rho_{e}$ are zero and hence in order to determine whether the frozen state has been reached in a particular realization, we focus on the time evolution of the link densities $\rho_i(t)$, \textit{$i\in$\{a, b, c, d, e, f\}}. This evolution has been analyzed, either using the Monte Carlo procedure explained before or by a set of approximate rate equations, derived in the Method section.

In the Erd\"os-R\'enyi Network, where links amongst nodes are generated randomly with probability $\mu/N$, it is possible to relate the different link densities with the number $n(t)$ of nodes holding the white opinion. The relations are
\begin{eqnarray}
\nonumber
\rho_{a}+\rho_{b}&\simeq&\frac{n(n-1)}{N(N-1)} \xrightarrow{n,N\gg1}x^2 \nonumber \\
\rho_{c}+\rho_{d}&\simeq&\frac{(N-n)(N-n-1)}{N(N-1)}\xrightarrow{n,N\gg1}(1-x)^2 \label{rho_rel}\\
\rho_{e}+\rho_{f}&\simeq&\frac{N(N-1)-(N-n)(N-n-1)-n(n-1)}{N(N-1)}\xrightarrow{n,N\gg1}2x(1-x)\nonumber
\end{eqnarray}
where $x=\dfrac{n}{N}$.

\section{Rate equation, fixed points and critical line}
As described in the Method section, we use a mean-field approximation to derive the rate equations of the aforementioned dynamics, based on the update rules sketched in Fig.~\ref{update_rule}. The main assumption in the derivation is that each node has exactly $\mu$ neighbors distributed randomly amongst all possible nodes. The mean-field treatment is exact in the thermodynamic limit (an infinite system size $N$) of an all-to-all network where all nodes are connected to each other, but it is only an approximation for other networks that we consider in our simulations. This is the case of an Erd\"os-R\'enyi network with a Poisson distribution, of mean degree $\mu$, for the number of neighbors. Furthermore, as the rate equations are deterministic, they can not describe the finite-size fluctuations observed in the numerical simulations. The rate equations are six coupled differential equations for the time evolution of the six types of pair densities $\{\rho_a,\rho_b,\rho_c,\rho_d,\rho_e,\rho_f\}$ and include a combination of linear and nonlinear terms of those variables, see Eqs. (\ref{eq:rate}). The linear terms reflect the direct change of densities in any type of update, either node or link, and the nonlinear terms are the indirect consequence of node update, by which the status of other links connected to the updated node will also change. The latter effect plays a crucial role in the evolution of the system. We first identify the fixed points of our 6th-order dynamical system, found by setting all time derivatives equal to zero. It turns out that there are two independent sets of fixed points. In the first set all densities take a well defined, non-null, value
\begin{eqnarray}
\rho^\text{st}_{a}=\frac{\rho^\text{st}_{e}}{2}, \quad \rho^\text{st}_{c}=\frac{\rho^\text{st}_{e}}{2}, \quad\rho^\text{st}_{e} = \frac{-3 + 2 (1-p)(\mu-1)}{8 (1-p)(\mu-1)},\quad \rho^\text{st}_{b}=\frac{\rho^\text{st}_{f}}{2},\quad \rho^\text{st}_{d}=\frac{\rho^\text{st}_{f}}{2},\quad \rho^\text{st}_{f} = \frac{1}{2}-\rho^\text{st}_e;
\label{first_set}
\end{eqnarray}
while in the second set
\begin{eqnarray}
\rho^\text{st}_{a} = 0,\quad \rho^\text{st}_{c}= 0,\quad \rho^\text{st}_{e}= 0,\quad (\rho^\text{st}_{b},\rho^\text{st}_{d},\rho^\text{st}_{f}) \rightarrow \textrm{arbitrary},
\label{second_set}
\end{eqnarray}
there are no unsatisfying pairs, but there is an arbitrariness in the values of the densities of the satisfying links (always verifying the normalization condition $\rho^\text{st}_b+\rho^\text{st}_d+\rho^\text{st}_f=1$).
In the first solution, Eq.(\ref{first_set}), the densities of the pair connections reach asymptotically a non-null plateau depending on $p$ and $\mu$ and independent of initial conditions. However, in the second set of solutions, Eq.(\ref{second_set}), there are no unsatisfying pair connections, and the densities of the satisfying ones depend on the initial conditions.

The condition $\rho^\text{st}_e\ge0$ determines that the first solution is only relevant for $-3+2(1-p)(\mu-1)\ge0$ or, given $\mu$, for $p\le p_{c}(\mu)$ with
\begin{equation}\label{eq:pc}
p_c(\mu)= 1-\frac{3}{2(\mu-1)}.
\end{equation}
A linear stability analysis\cite{strogatz2018nonlinear} shows that the first solution, Eq.(\ref{first_set}), is always stable (negative eigenvalues of the linearized equations) whenever it leads to non-negative densities, and that the set, Eq.(\ref{second_set}), is marginally stable (eigenvalues equal to zero). Therefore for $p<p_c(\mu)$ there are always active links in the asymptotic state and, consequently, continuous changes of the microscopic state. We say that the system stays in an {\bf active}, or dynamical, phase. However, for $p>p_c(\mu)$, the densities of all active pairs tend to zero and the dynamics reaches a {\bf frozen}, or absorbing, phase. As order parameter distinguishing one phase from another we choose the density $\rho^\text{st}_e$, which is zero in the frozen phase and positive in the active phase. Note that, in turn, the condition $p_c(\mu)\ge0$ requires $\mu>5/2$. If $\mu\le 5/2$ the first solution does not exist and the dynamics always leads to a frozen phase. As $p$ approaches the critical value $p_c$ from below, the order parameter tends to zero as $\rho^\text{st}_e\sim(p_c-p)^\beta$ with a critical exponent $\beta=1$, a continuous phase transition. It is known that the critical exponent of the directed percolation for mean field and in spatial dimension larger than 4 is $\beta=1$. Thus, the active-frozen transition that we just described can be categorized under the class of directed percolation which is a very important class of absorbing transitions in the non-equilibrium critical phenomena\cite{hinrichsen2000non}.

These conclusions are based on the study of the rate equations and are strictly valid only in the thermodynamic limit for the all-to-all network. In a finite system, the active phase will display fluctuations of the order parameter around its mean value. Due to the stochastic nature of the dynamics, there will be always a fluctuation that takes the active phase into the frozen one and, from there on, all microscopic dynamics stops. As discussed in detail in the next sections, the likeness of such fluctuation tends to zero with increasing system size and the average time to reach the frozen phase diverges exponentially with system size.

\section{Phase Transition}

We have carried out extensive Monte-Carlo simulations of the dynamical rules presented in Fig.~\ref{update_rule} on an Erd\"os-R\'enyi network with mean degree $\mu$ and system size $N$. Once the network, nodes and links, has been constructed, we assign randomly a state (white/blue) to each node and then a state to each link (friendly/unfriendly). The initial density of white nodes is $x_0$ and that of friendly links is $\ell_0$. Note that the all-to-all network corresponds to $\mu=N-1$.

\begin{figure}[]
 \subfigure[]{
 \includegraphics[width=0.48\textwidth]{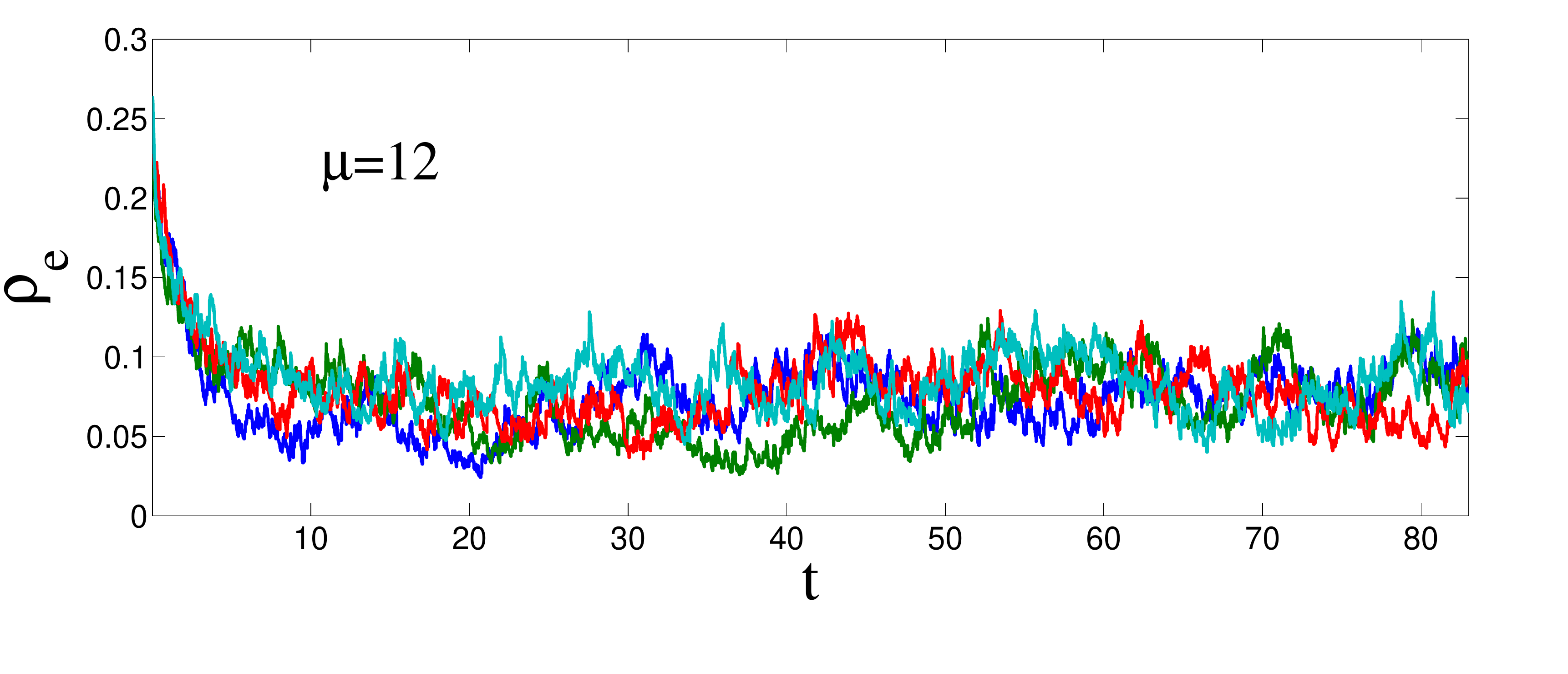}
}
 \subfigure[]{
 \includegraphics[width=0.48\textwidth]{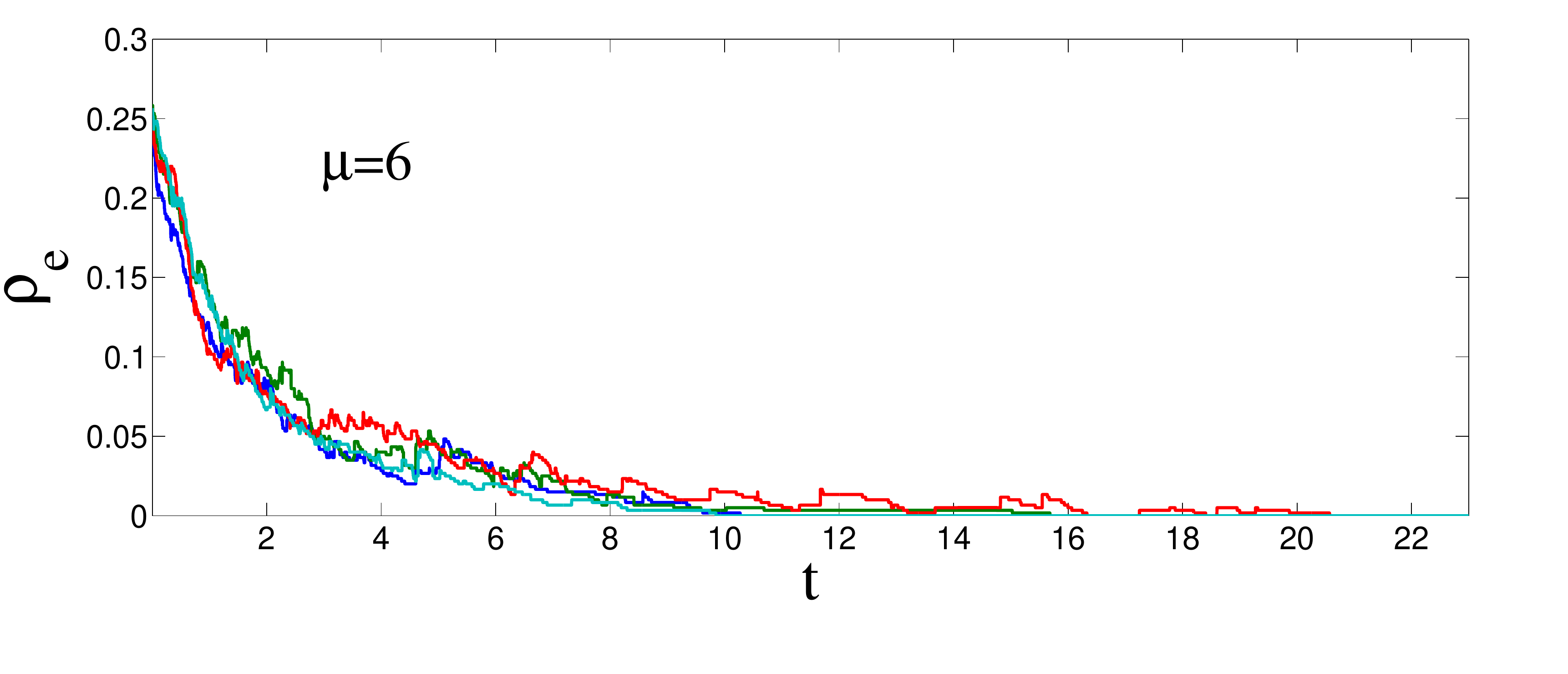}
}
\caption{We plot some representative trajectories of the density $\rho_{e}(t)$ of unsatisfied links, for the Erd\"os-R\'enyi network with $N=200$. The relevant parameters for panel (a) are $p=0.8$ and $\mu=12$, leading to an active phase where the evolution continues for the whole range of time displayed. In panel (b) we take $p=0.8$ and $\mu=6$, leading to a frozen phase with a zero density of unsatisfying links. All trajectories are generated using the Monte Carlo method described in the main text and, in both panels, start out from the same value of the density of white nodes, $x_0=0.5$, and the density of friendly links, $\ell_0=0.5$, but correspond to different realizations of the network, initial condition and dynamics.}
\label{dynamics_single}
\end{figure}

Some representative trajectories of the order parameter $\rho_e(t)$ can be seen in Fig.~\ref{dynamics_single} for $N=200$ for some system parameters leading to the active phase, panel (a), or to the frozen phase, panel (b). In all curves we have taken the same values of $x_0=0.5$ and $\ell_0=0.5$ but different realizations of the networks, initial conditions and the dynamics. Observe the dispersion in the different curves in the active phase and that the microscopic dynamics continues in this phase for the whole range of time displayed in the figure. In the frozen phase, there is no further dynamics when the density of unsatisfied links reaches zero. Note, however, that there is also a dispersion in the times it takes the different realizations of the dynamics to reach the frozen state.

\begin{figure}[]
\centering
 \subfigure[]{
 \centering
 \includegraphics[width=0.45\textwidth]{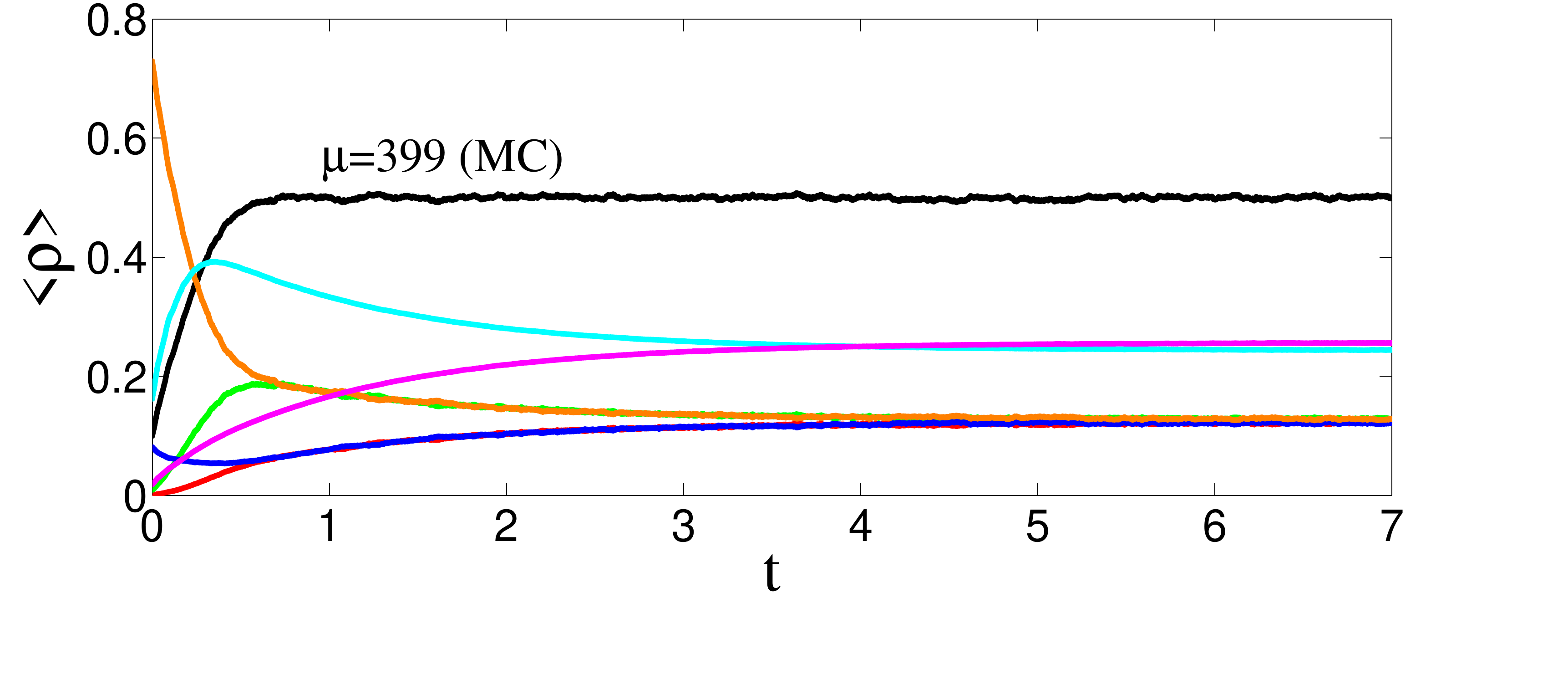}
}
 \subfigure[]{
 \centering
 \includegraphics[width=0.45\textwidth]{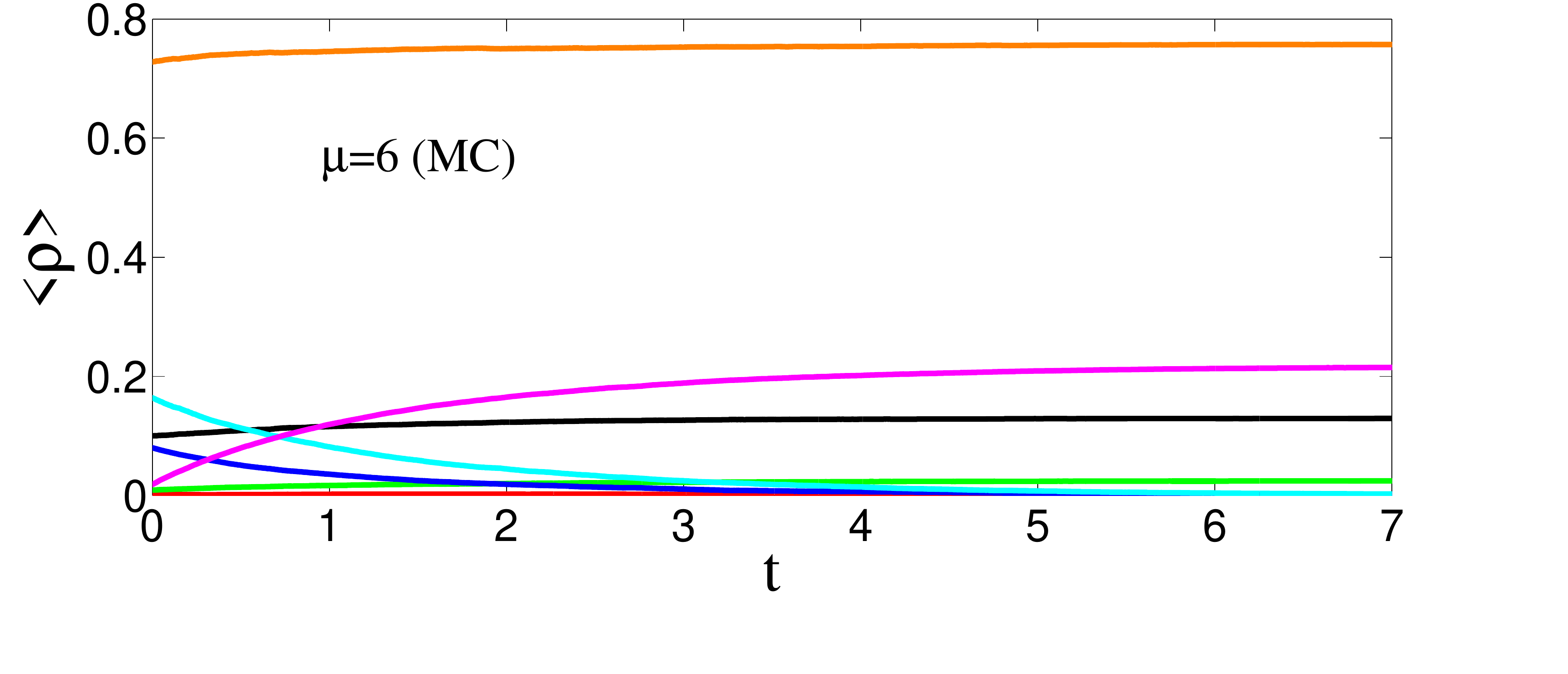}
}
 \subfigure[]{
 \centering
 \includegraphics[width=0.45\textwidth]{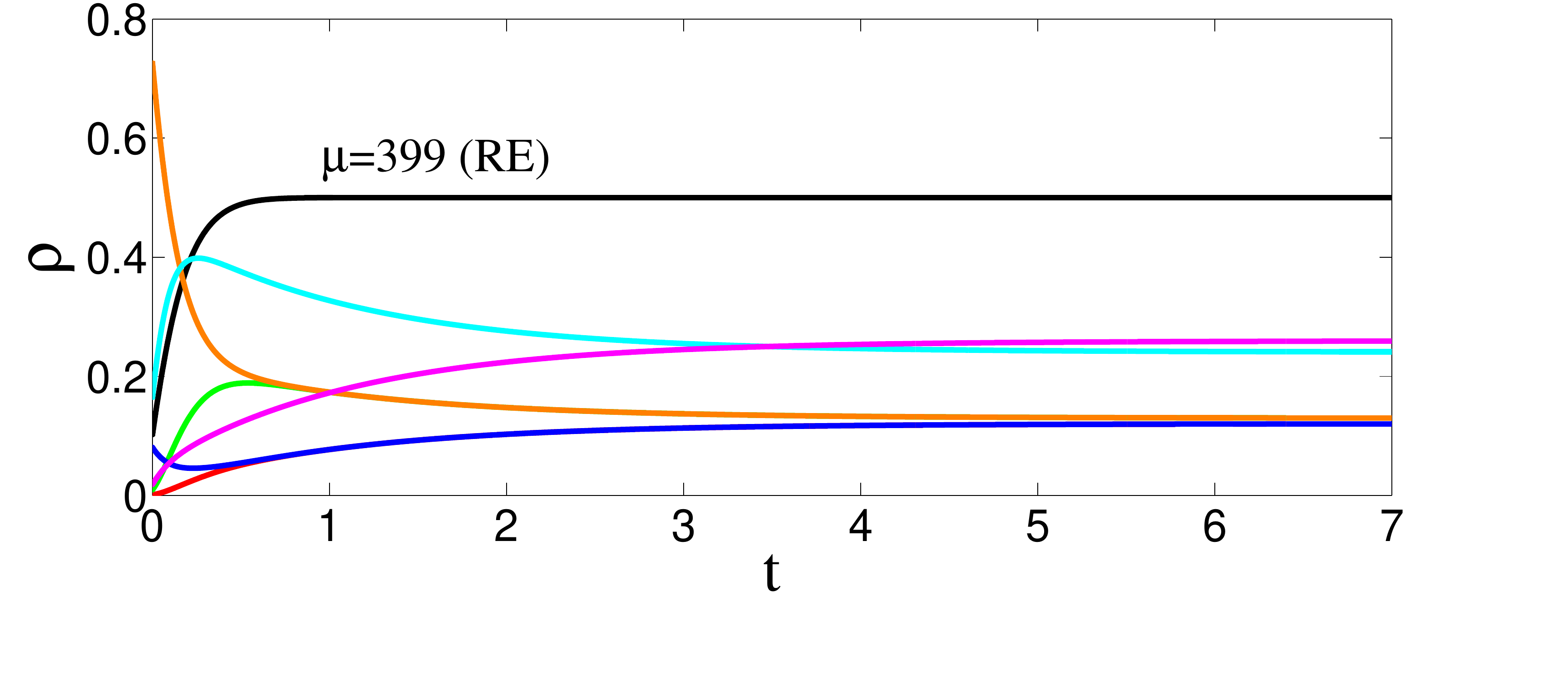}
}
 \subfigure[]{
 \centering
 \includegraphics[width=0.45\textwidth]{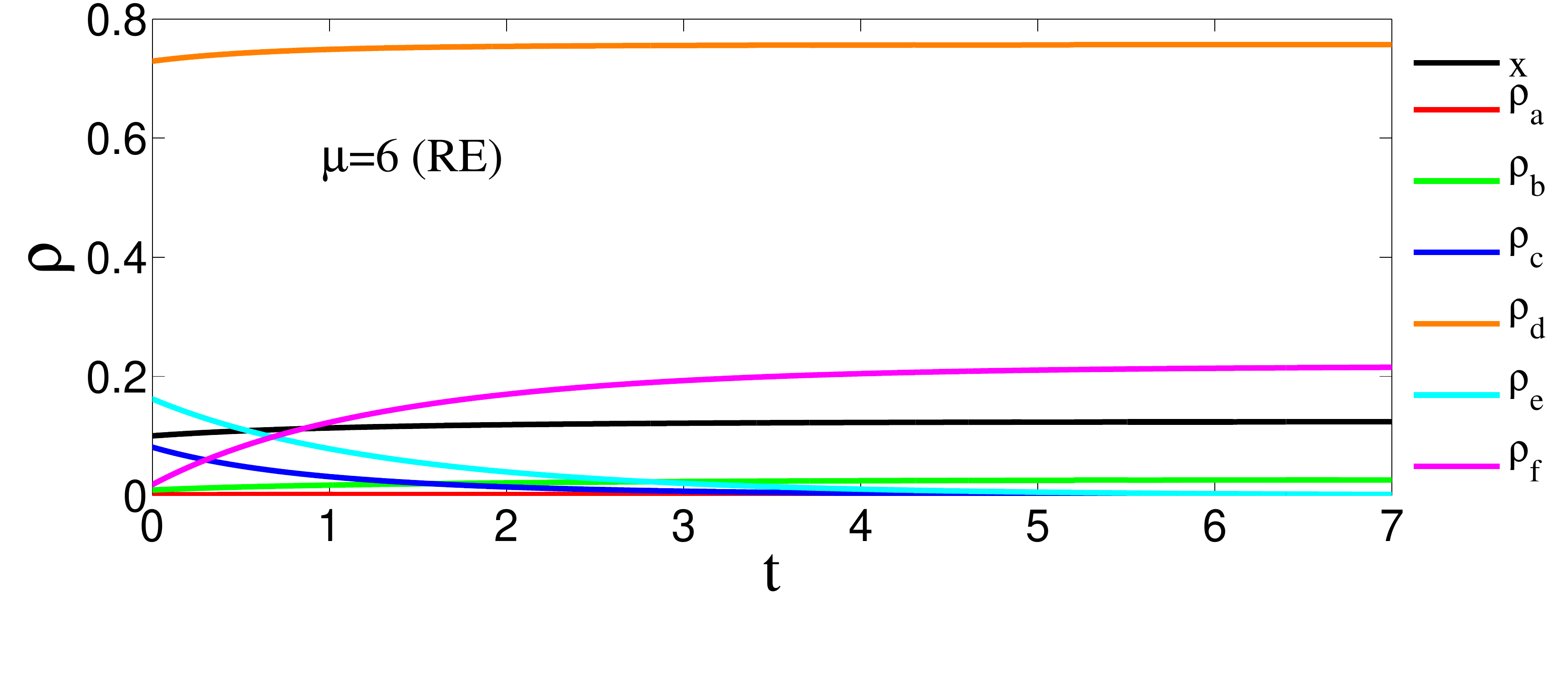}
}
\caption{Densities of the different types of links $\{\rho_a(t),\rho_b(t),\rho_c(t),\rho_d(t),\rho_e(t),\rho_f(t)\}$ and the density of white nodes $x(t)$ as a function of time. Panels (c) and (d) are the result of a numerical integration of the rate equations (RE), and have to be compared with the results of the Monte Carlo (MC) simulations with the same value of $\mu$, displayed in panels (a) and (b). All curves use $p=0.9$ and start out from same value, $x_0=0.1$ and $\ell_0=0.9$. The curves obtained from the Monte Carlo simulation are the result on an average over 100 simulation runs on the Erd\"os-R\'enyi network with $N=400$. Panels (a) and (c) correspond to the active phase while panels (b) and (d) correspond to the frozen phase.}
\label{dynamics}
\end{figure}

In order to check the validity of the description in terms of rate equations, we compare in Fig.~\ref{dynamics} the Monte-Carlo results for the time evolution of the densities $\{\rho_a,\rho_b,\rho_c,\rho_d,\rho_d,\rho_f\}$ for an Erd\"os-R\'enyi network with $N=400$ nodes, with the numerical integration of the rate equations. We have set the same average number of neighbors $\mu$ in the rate equations and the Monte-Carlo simulations. For the Monte-Carlo simulations the results are the average over $100$ realizations. In panels (a) and (c) we take $\mu=399$ which, for the Erd\"os-R\'enyi network, means that all nodes are connected to each other. In this case, as expected, the agreement between the simulations and the rate equations is very good. In panels (b) and (d) we take $\mu=6$, but still observe a good agreement between simulations and rate equations. For $\mu=399$, the chosen value of $p=0.9$ is below the critical one $p<p_c(\mu)=0.996$ and, consequently, the dynamics leads to the active phase. For $\mu=6$, on the contrary, the value of $p=0.9$ is above the critical one $p>p_c(\mu)=0.7$ and, consequently, the dynamics leads to the frozen phase. These two scenarios are observed both in the Monte-Carlo simulations and in the numerical integration of the rate equations.

\begin{figure}[]
\centering
 \subfigure[]{
 \centering
 \includegraphics[width=0.28\textwidth]{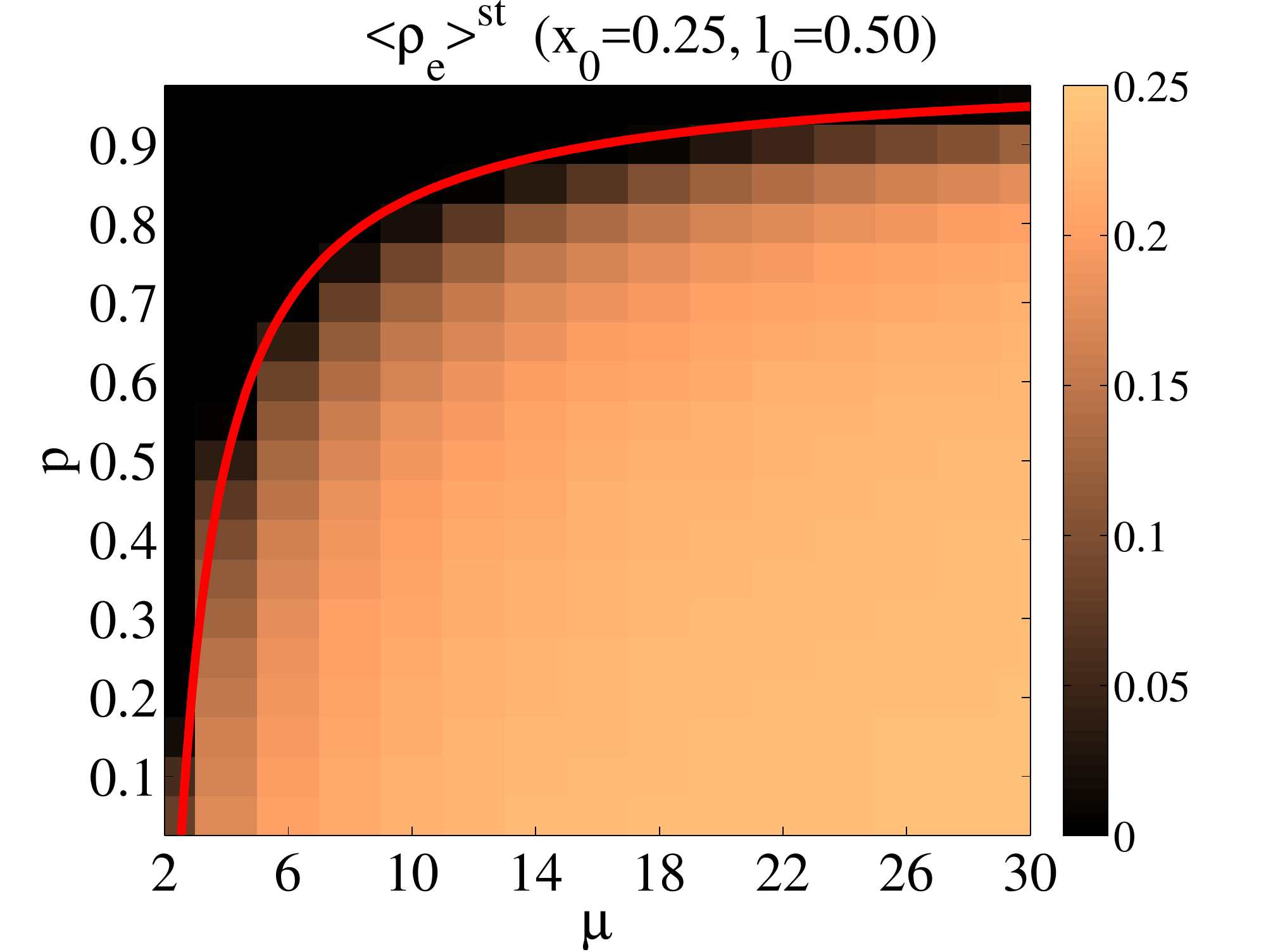}
}
 \subfigure[]{
 \centering
 \includegraphics[width=0.285\textwidth]{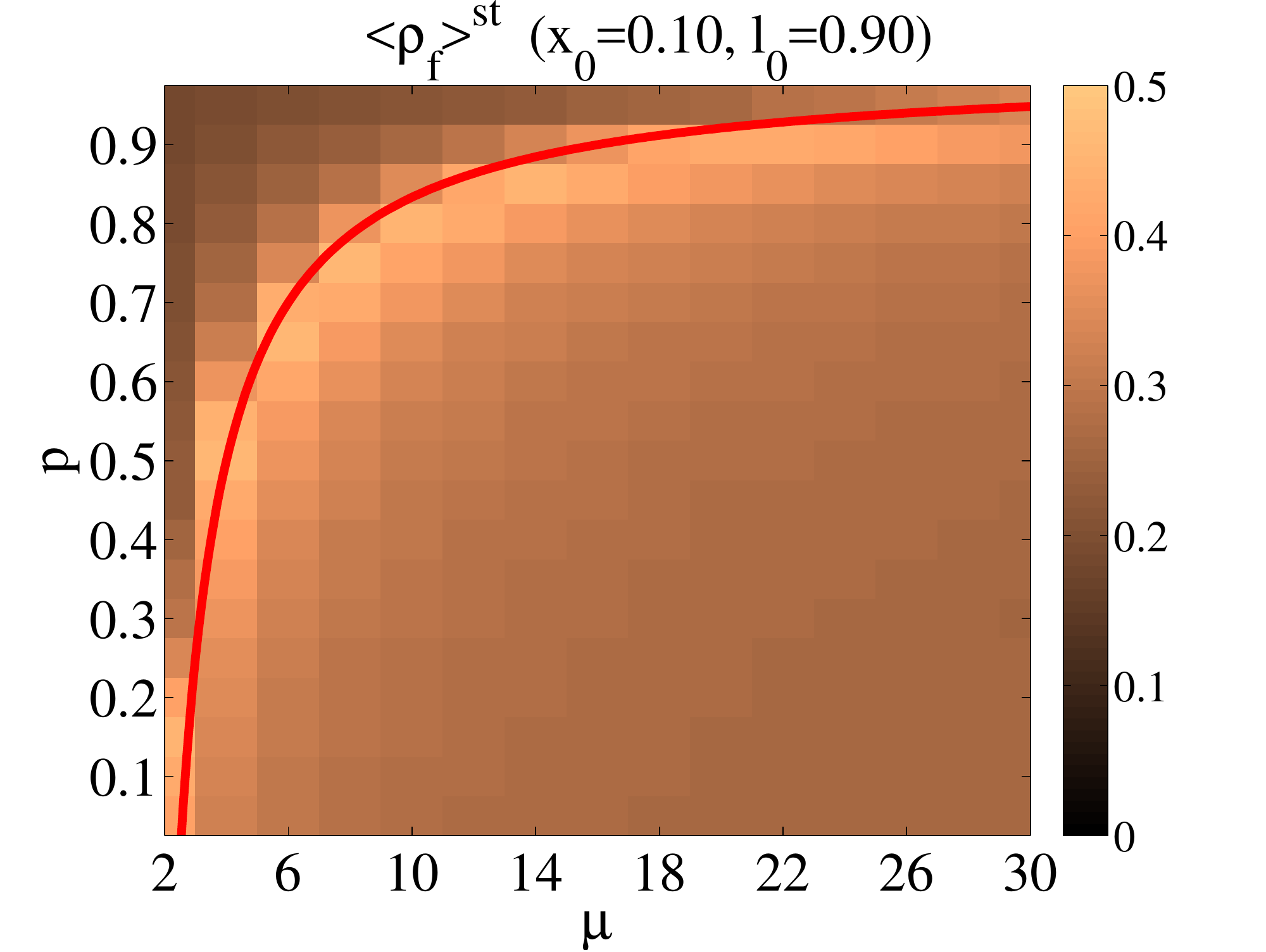}
}
 \subfigure[]{
 \centering
 \includegraphics[width=0.285\textwidth]{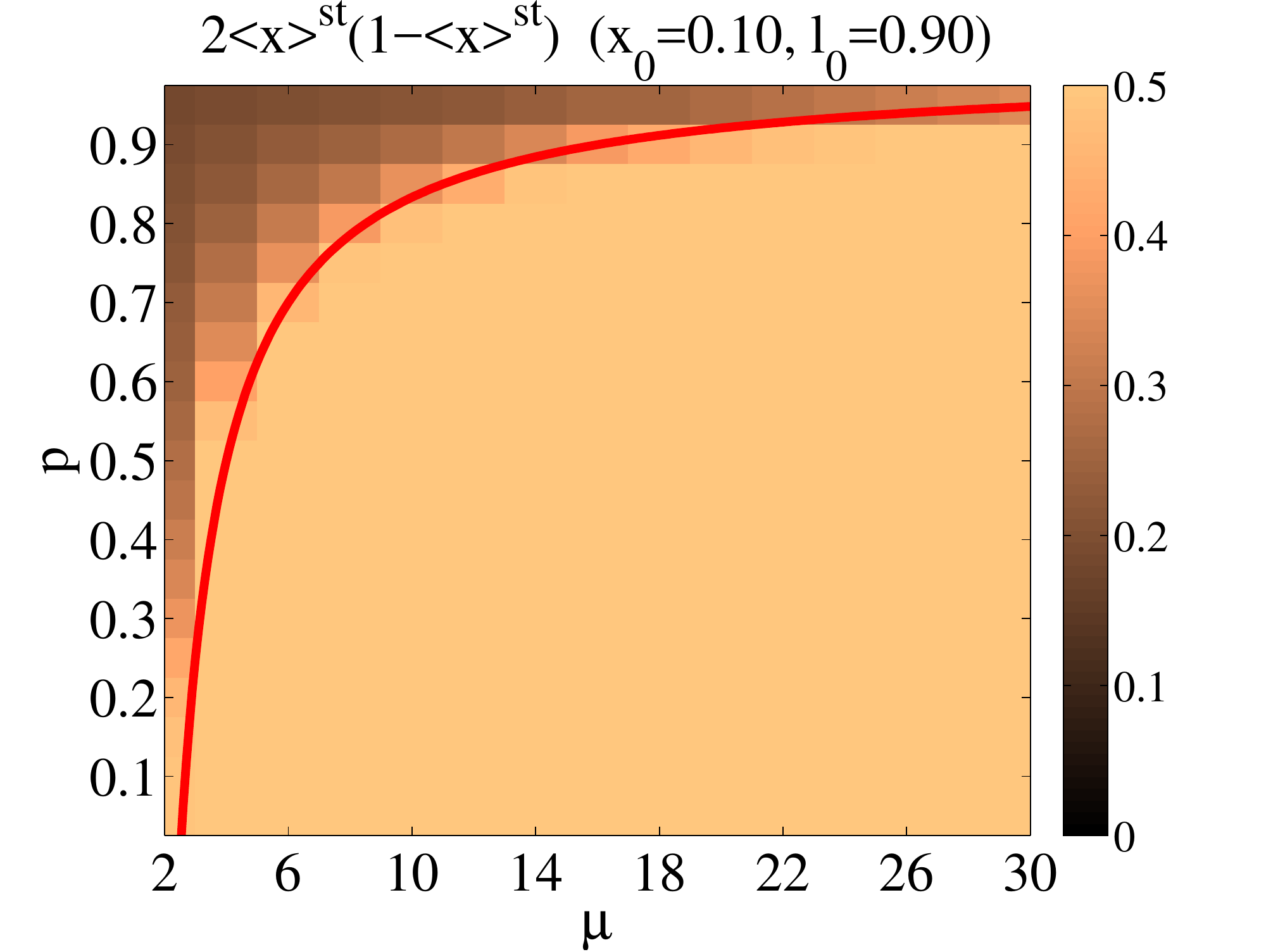}
}
 \subfigure[]{
 \centering
 \includegraphics[width=0.285\textwidth]{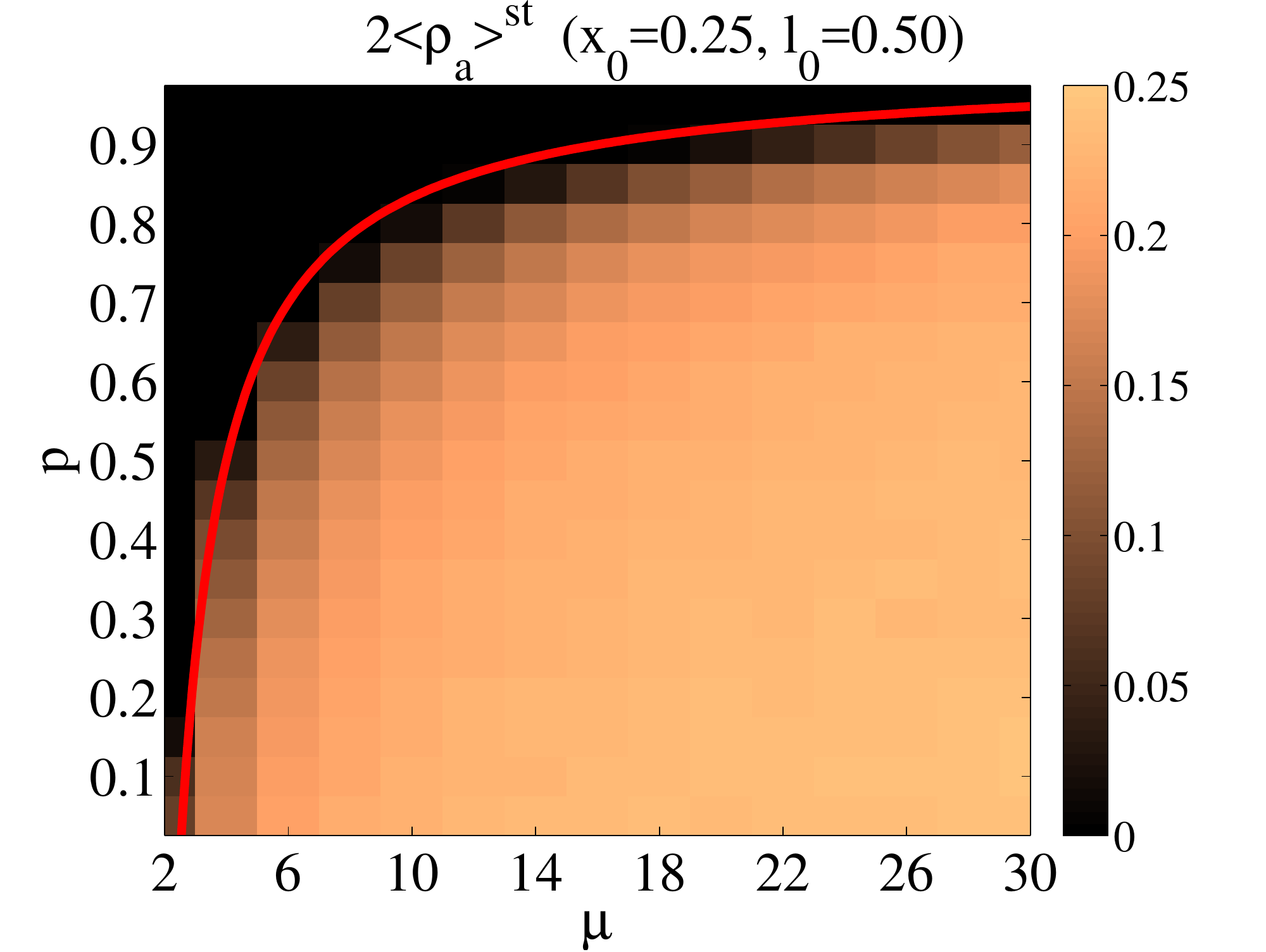}
}
 \subfigure[]{
 \centering
 \includegraphics[width=0.285\textwidth]{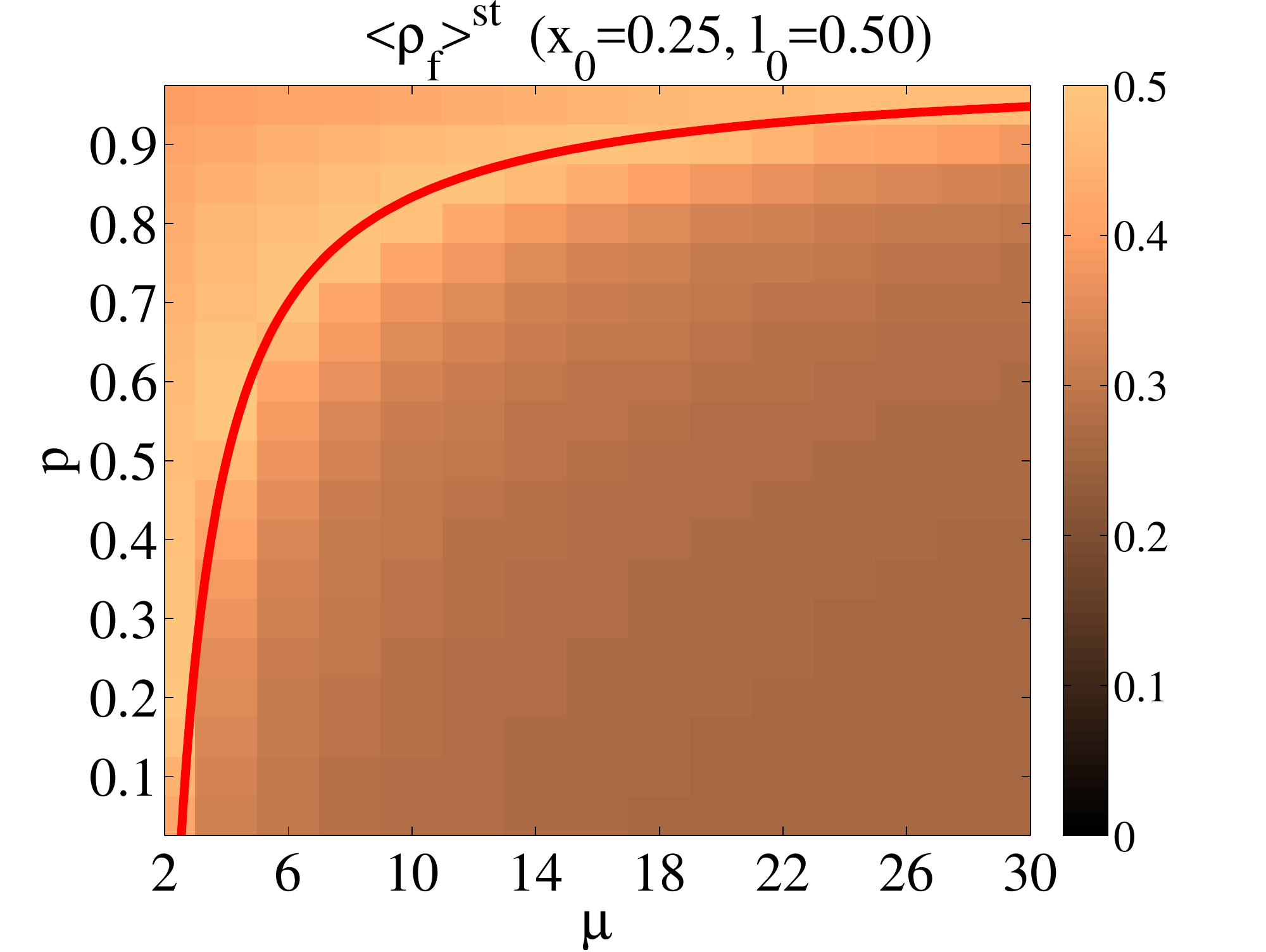}
}
 \subfigure[]{
 \centering
 \includegraphics[width=0.285\textwidth]{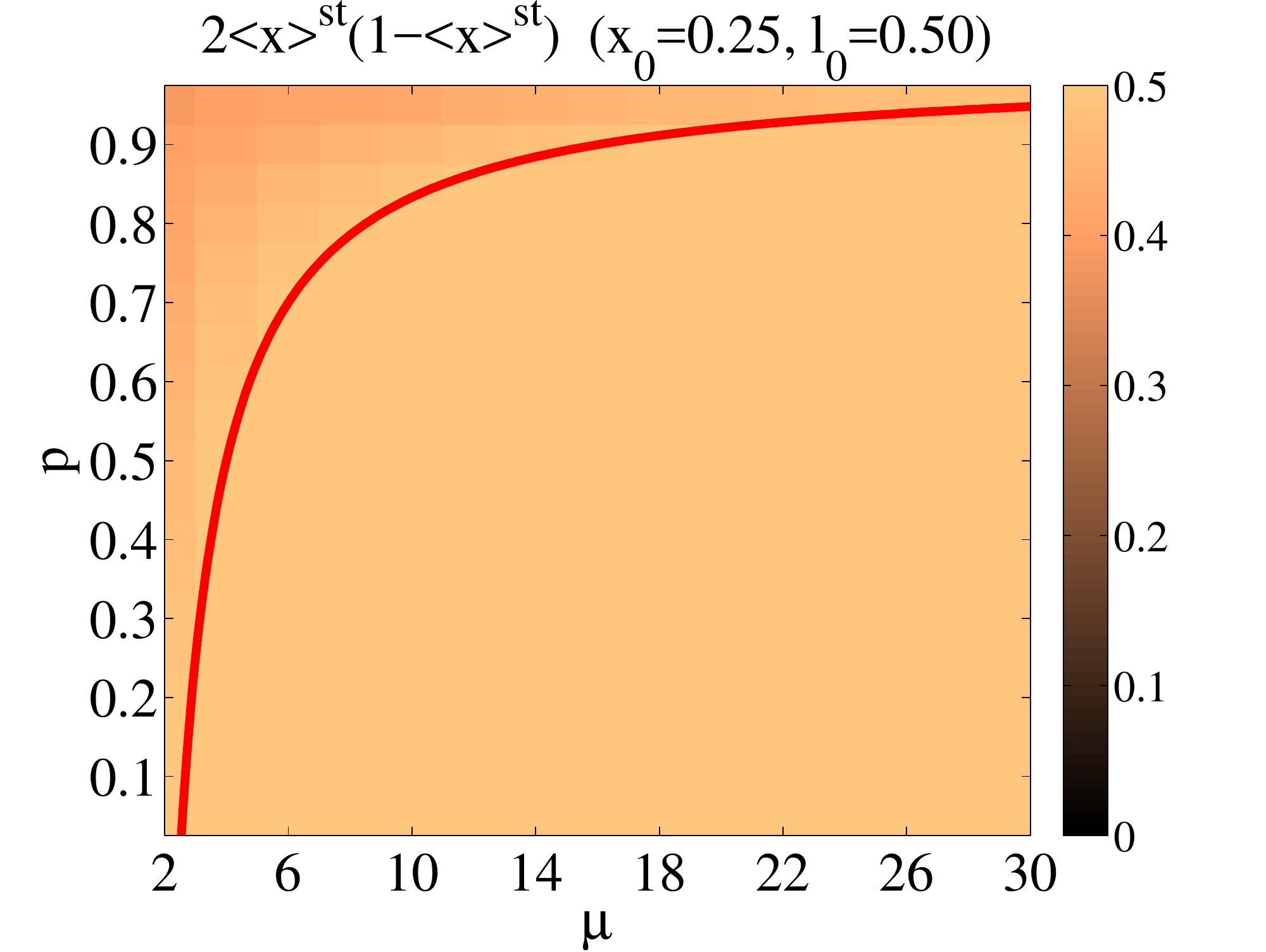}
}
\caption{These figures plot in a color code the density of links obtained from Monte Carlo simulation as a function of $\mu$ and $p$. The critical line obtained from the rate equations is plotted as a red line. As it can be seen, the values of $\langle\rho_{e}\rangle^{\text{st}}$ displayed in panel (a) and $2\langle\rho_{a}\rangle^{\text{st}}$ in panel (d) are indistinguishable as it was predicted by Eqs.(\ref{first_set},\ref{second_set}), $\rho_{a}=\rho_{e}/2$ in active phase and $\rho_{a}=\rho_{e}=0$ in frozen phase. Panels (b) and (e) represent $\rho_{f}$ for different initial conditions, confirming that in the frozen phase (above the critical line) the density of unsatisfying pairs depends on the initial condition, while in the active phase (below the critical line) they are independent of the initial condition. Finally, based on Eqs.(\ref{rho_rel}), the rate equations predict that the relation $\rho_f^\text{st}=2x^\text{st}(1-x^\text{st})$ should hold in the frozen phase as it is indeed the case: compare, above the critical line, panels (c) and (b) for one initial condition and panels (e) and (f) for a different one.}
\label{PT}
\end{figure}

\begin{figure}[]
\centering
 \subfigure[]{
 \centering
 \includegraphics[width=0.34\textwidth]{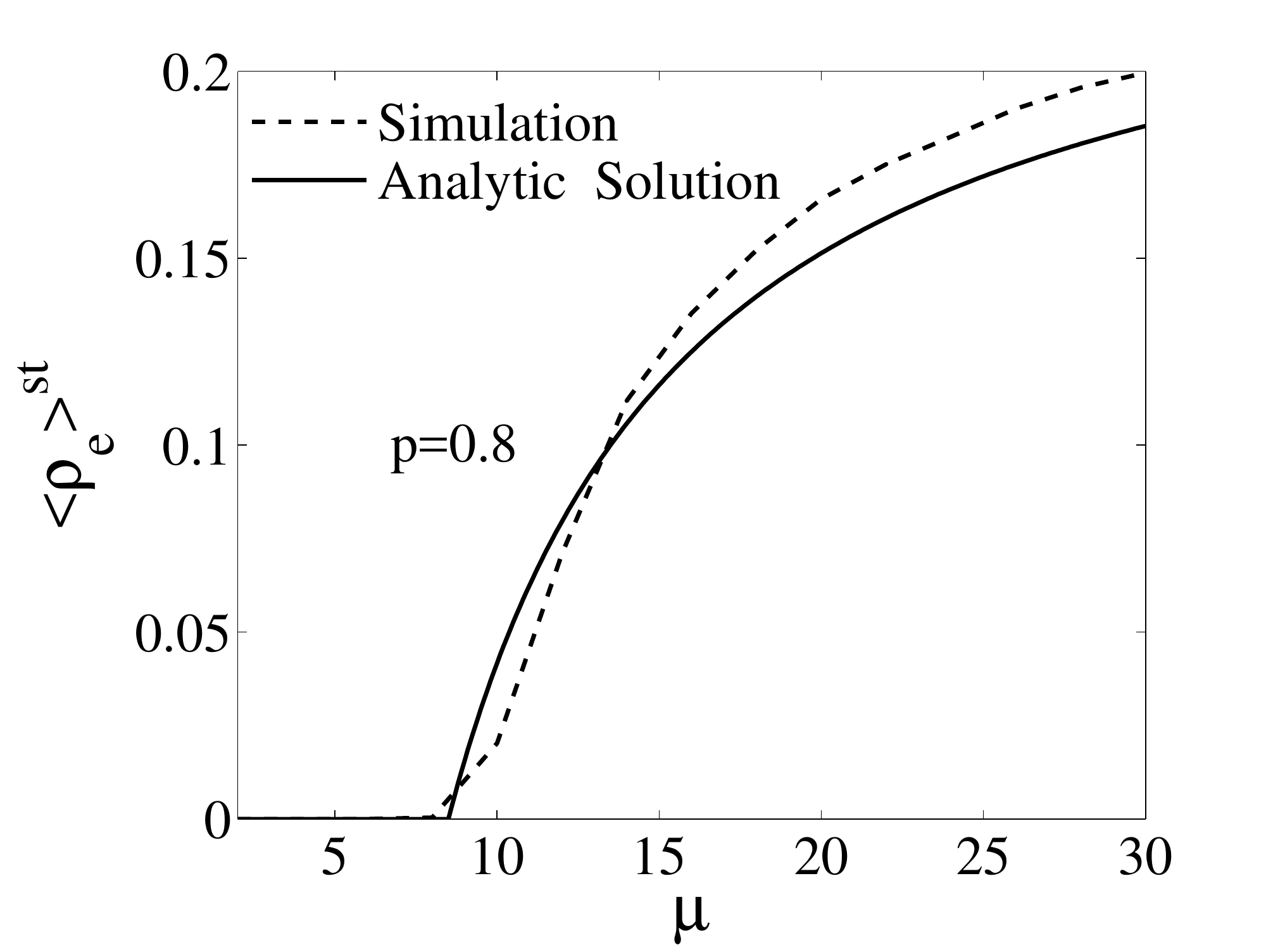}
}
 \subfigure[]{
 \centering
 \includegraphics[width=0.34\textwidth]{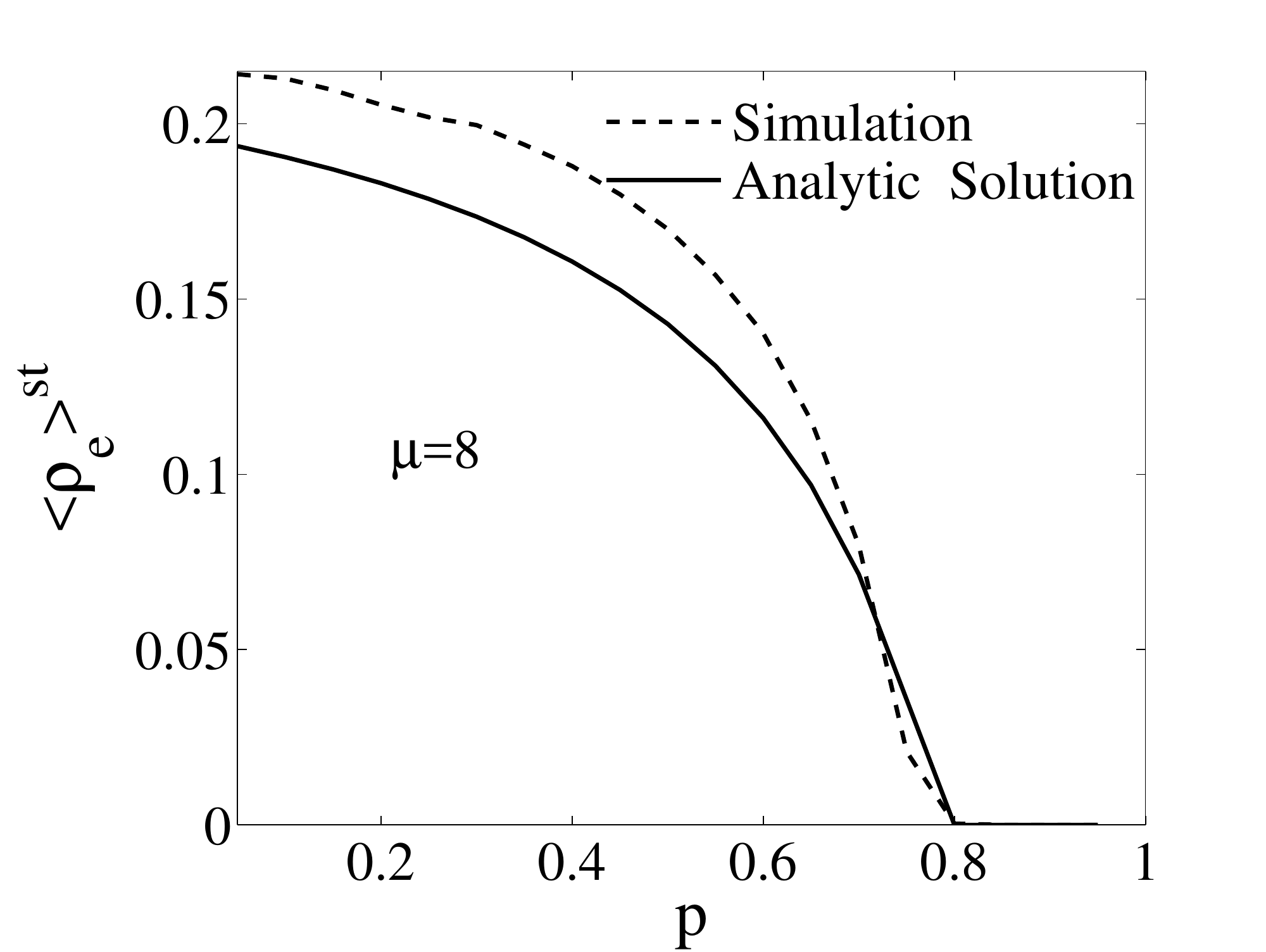}
}
\caption{We plot the stationary density average $\langle\rho_{e}\rangle^{\text{st}}$ obtained from Monte Carlo simulations (dashed lines) and compare it with the analytical solution of the rate equations (solid lines). In panel (a) we plot this magnitude as a function of $\mu$ for fixed $p=0.8$, while in panel (b) we plot it as function of $p$ for fixed $\mu=8$. }
\label{lattice1}
\end{figure}

To compare the Monte-Carlo simulations with the predictions of the rate equations in the steady state, Eqs.(\ref{first_set},\ref{second_set}) and the critical line Eq.(\ref{eq:pc}), we compute numerically in the simulations the average value $\langle \rho_e\rangle^\text{st}$. In Fig.~\ref{PT}(a) we show $\langle \rho_e\rangle^\text{st}$ in the $(\mu,p)$ plane in a color code, while in Fig.~\ref{lattice1}(a) we plot it as a function of $\mu$ for fixed $p=0.8$ and in Fig.~\ref{lattice1}(b) as a function of $p$ for fixed $\mu=8$, as well as the prediction of the rate equations. It is clear from these figures that the rate equations provide a good qualitative, but also quantitative, agreement with the results of the Monte-Carlo simulations in finite Erd\"os-R\'enyi lattices. The rate equations also predict the relation $\rho^\text{st}_a=\rho^\text{st}_e/2$ that occurs in the steady state. This has been checked by plotting in Fig.~\ref{PT}(d), $2\langle \rho_a\rangle^\text{st}$ in the $(\mu,p)$ plane in a color code, making it indistinguishable from the corresponding Fig.~\ref{PT}(a) for $\langle \rho_e\rangle^\text{st}$. Another prediction of the rate equations is that $\rho^\text{st}_f$ should depend on the initial condition in the frozen phase. We check this by plotting $\langle \rho_f\rangle^\text{st}$ in the $(\mu,p)$ plane in a color code for two different values of the initial condition in Figs.\ref{PT}(b) and \ref{PT}(e).

For an Erd\"os-R\'enyi network, Eqs.(\ref{rho_rel}-\ref{second_set}) imply that the density of white links in the steady state should be $x^\text{st}=1/2$ in the active phase, while it depends on the initial condition in the frozen phase. Moreover, in the frozen phase, the relation $\rho_f^\text{st}=2x^\text{st}(1-x^\text{st})$ should hold. This is checked in Figs.\ref{PT}(c) and \ref{PT}(f) where we plot $2\langle x\rangle^\text{st}(1-\langle x\rangle^\text{st})$ in the $(\mu,p)$ plane in a color code yielding plots which are indistinguishable from the corresponding Figs.\ref{PT}(b) and \ref{PT}(e), respectively, for $\langle \rho_f\rangle^\text{st}$ in the frozen phase, $p>p_c(\mu)$.

\begin{figure}[]
\centering
 \subfigure[]{
 \centering
 \includegraphics[width=0.34\textwidth]{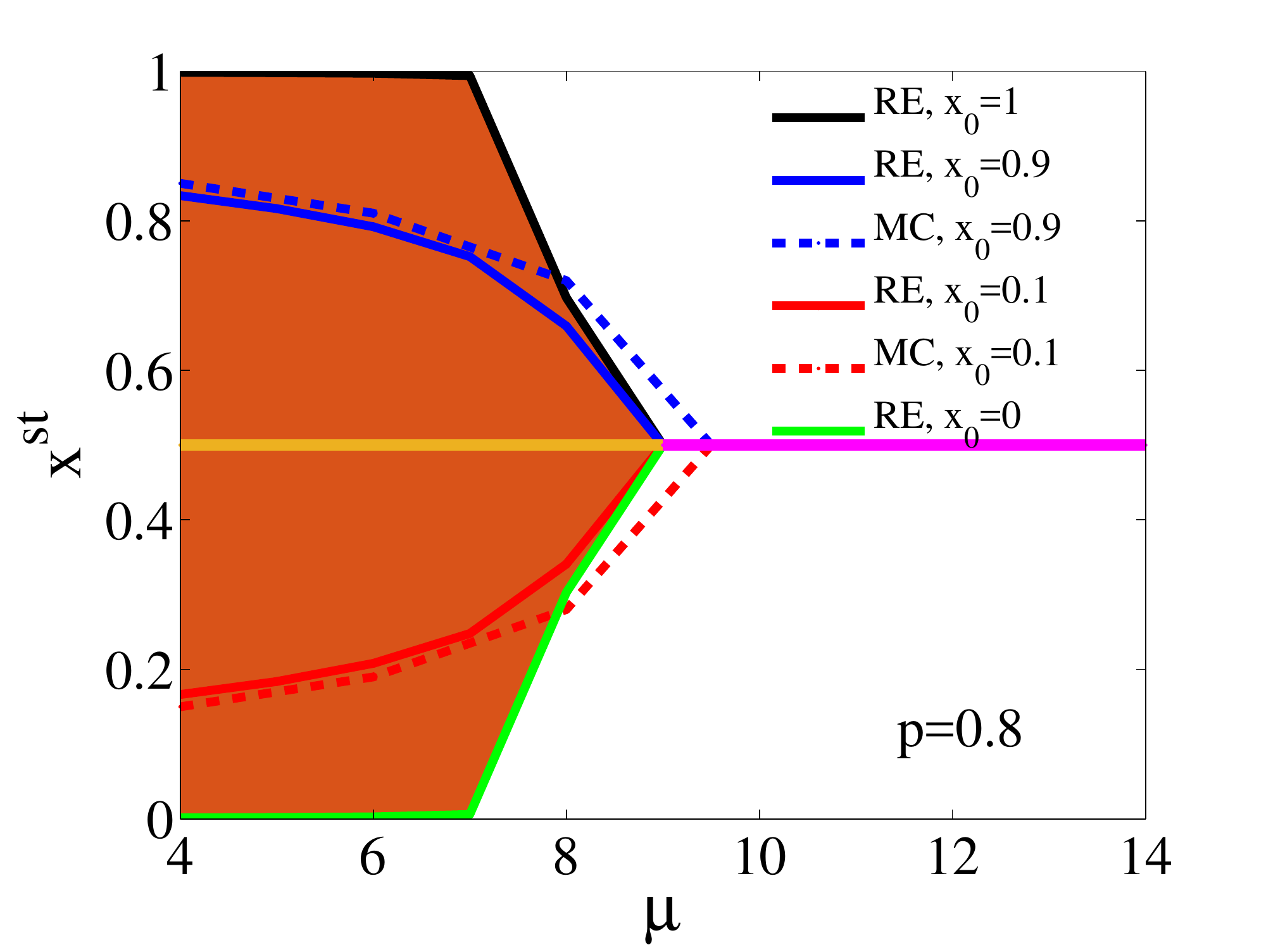}
}
 \subfigure[]{
 \centering
 \includegraphics[width=0.34\textwidth]{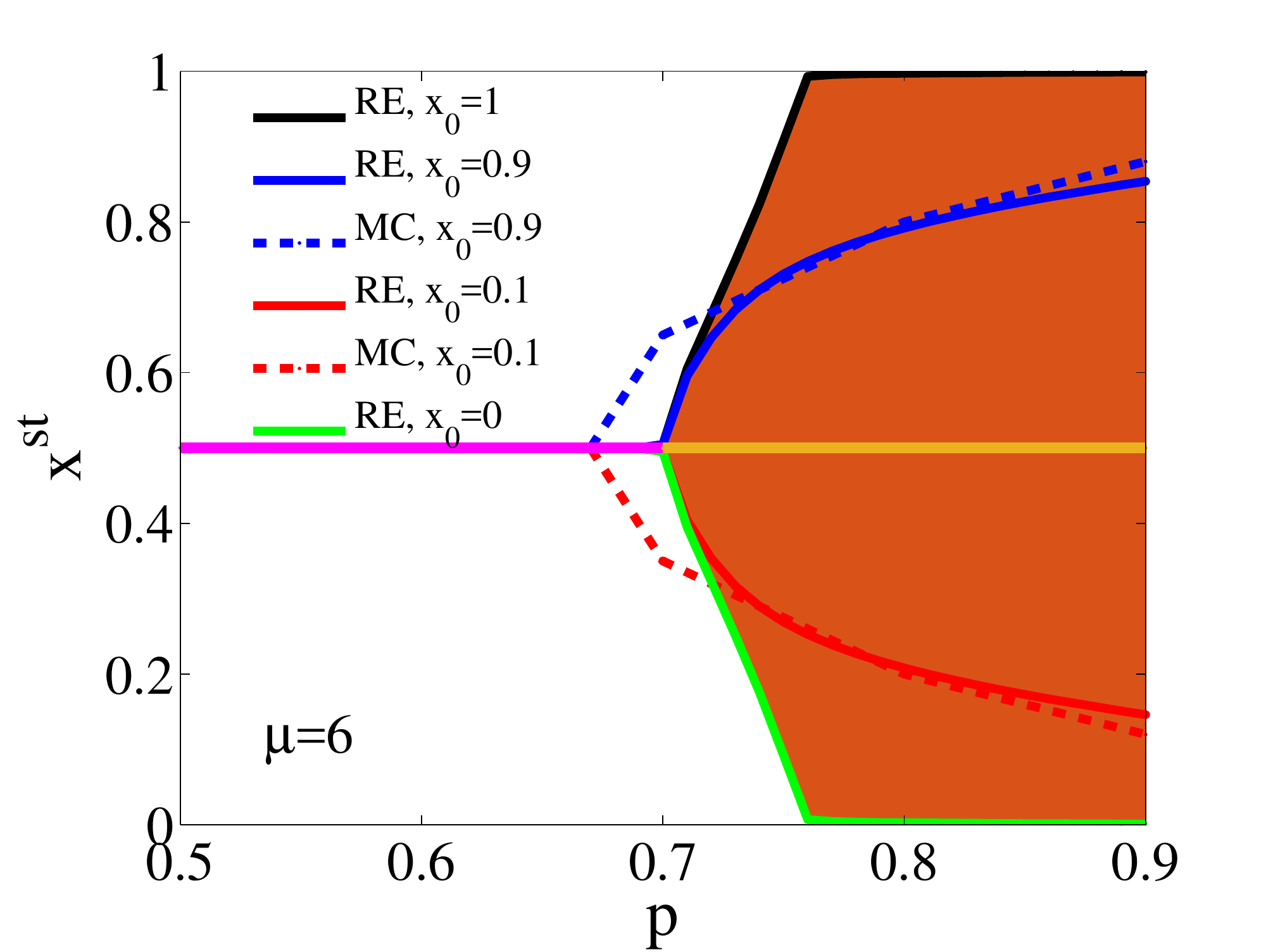}
}
\caption{We plot the dependence of $x^\text{st}$ on the initial condition in the frozen phase both for fixed $p=0.8$, panel (c) and fixed $\mu=6$, panel (d). In the frozen phase, the spread of $x^\text{st}$, due to varying $\ell_0$ from $0$ to $1$, for given $x_0$ are spanned between the $x_f$ (curves) and $x=1/2$ (beige line). The colored area is the region between the largest and smallest possible value of $x_f$ for which we used the condition $\ell_0=\ell_1=1-\ell_2$ (see initial condition in section method). In active phase, $p<p_c(\mu)$, for any initial condition, $x^\text{st}=1/2$ (pink line). Finally, for comparison, $x_f$ obtained from Monte Carlo for $x_0=0.1,0.9$ are depicted by dash line.}
\label{lattice2}
\end{figure}

A more detailed analysis of the dependence of $x^\text{st}$ and hence of $\rho^\text{st}_b$, $\rho^\text{st}_d$ and $\rho^\text{st}_f$ on the initial condition, in the frozen phase has been performed in Figs.~\ref{lattice2}(c) and ~\ref{lattice2}(d). We set a value of $x_0$ (initial density of white nodes) and generate many initial conditions varying the value of the initial density of friendly links $\ell_0\in(0,1)$. Each of these microscopic configurations evolves to $x^\text{st}=1/2$ for $p<p_c(\mu)$ but generates a spread of final values of $x^\text{st}$ for $p>p_c(\mu)$. The spread is delimited by the lines $x=1/2$ and $x=x_f$, being $x_f$ a value that depends on $x_0$, $\mu$ and $p$. The value $x_f(x_0,\mu)$ for fixed $p=0.8$ and the value $x_f(x_0,p)$ for fixed $\mu=6$ have been plotted in Figs.~\ref{lattice2}(c) and ~\ref{lattice2}(d), respectively, for different values of $x_0$.

We have also carried out numerical simulations in regular lattices in one and two dimensions with nearest neighbors (results not shown). The qualitative phenomenology is the same than the one described previously: while the dynamics always leads to a frozen phase in dimension one ($\mu=2$), in dimension two we still find a transition from an active to a frozen phase at a critical value $p_c$ that, however, is smaller than the equivalent $p_c(\mu)$ of an Erd\"os-R\'enyi network with $\mu=4$.

In the following sections we describe separately the active and the frozen phases, studying in each case the survival probability and the characteristic decay time, as well as some topological properties of the asymptotic pattern reached by the dynamical evolution.

\section{Active Phase}
The active, or dynamical, phase is a steady state characterized by the existence of a microscopic evolution. This occurs if the link-update probability is smaller than the critical value, $p<p_c(\mu)$. All types of links $\{a,b,c,d,e,f\}$ are present in this active steady state and their corresponding densities fluctuate around well defined values. The amplitude of these fluctuations around the steady state values decrease with system size and, eventually, tend to zero in the thermodynamic limit when the number of nodes $N\to\infty$. See Fig.~\ref{rho_e}(a) for a typical dependence of the density $\rho_e(t)$ with time and system size. As shown in the same figure, the steady-state value $\langle \rho_e\rangle^\text{st}$ can be well approximated by the prediction Eq.~(\ref{first_set}) of the rate equations.

The existence of absorbing states and the ergodicity of the stochastic dynamical rules imply that for a finite system there is always a fluctuation that will take the system towards one of the absorbing states. Therefore, for a finite system, the ultimate fate is to end in the absorbing phase. The key point is to analyze the dependence of the time $\tau$ to reach the absorbing state on system size. $\tau$ is a random variable and we present in panels (a) and (b) of Fig.~\ref{tau} its probability density function (pdf) $f(\tau)$ for two values of $(\mu,p)$ corresponding to the active phase and several values of the system size $N$. It appears from these figures that, at least for large $\tau$, the pdf can be fitted by an exponential form $f(\tau)\sim e^{-\tau/\langle\tau\rangle}$. The average value of the distribution $\langle \tau\rangle$ is plotted in panels (e) and (f) Fig.~\ref{tau} as a function of the system size $N$, showing an exponential dependence $\langle \tau\rangle\sim e^{\alpha N}$. This exponential dependence indicates that the decay to the absorbing state becomes very rare for increasing system size. From the point of view of Statistical Mechanics, in the limit $N\to\infty$, the active state remains forever and represents a genuine macroscopic phase.

The active phase is remarkable in the sense that it indicates the failure of the local dynamical rule. While the evolution is dictated by a tendency to reduce the unsatisfying pairs, the final state is one of coexistence of all types of links. This is reminiscent of other dynamical models, the most notable being that of Axelrod\cite{axelrod1997dissemination,Axelrod2} that predicts that local convergence can generate global polarization in an agent based model of dissemination of culture.

\begin{figure}[]
\centering
 \subfigure[]{
 \centering
 \includegraphics[clip, trim=0cm 15cm 0cm 0cm, width=0.48\textwidth]{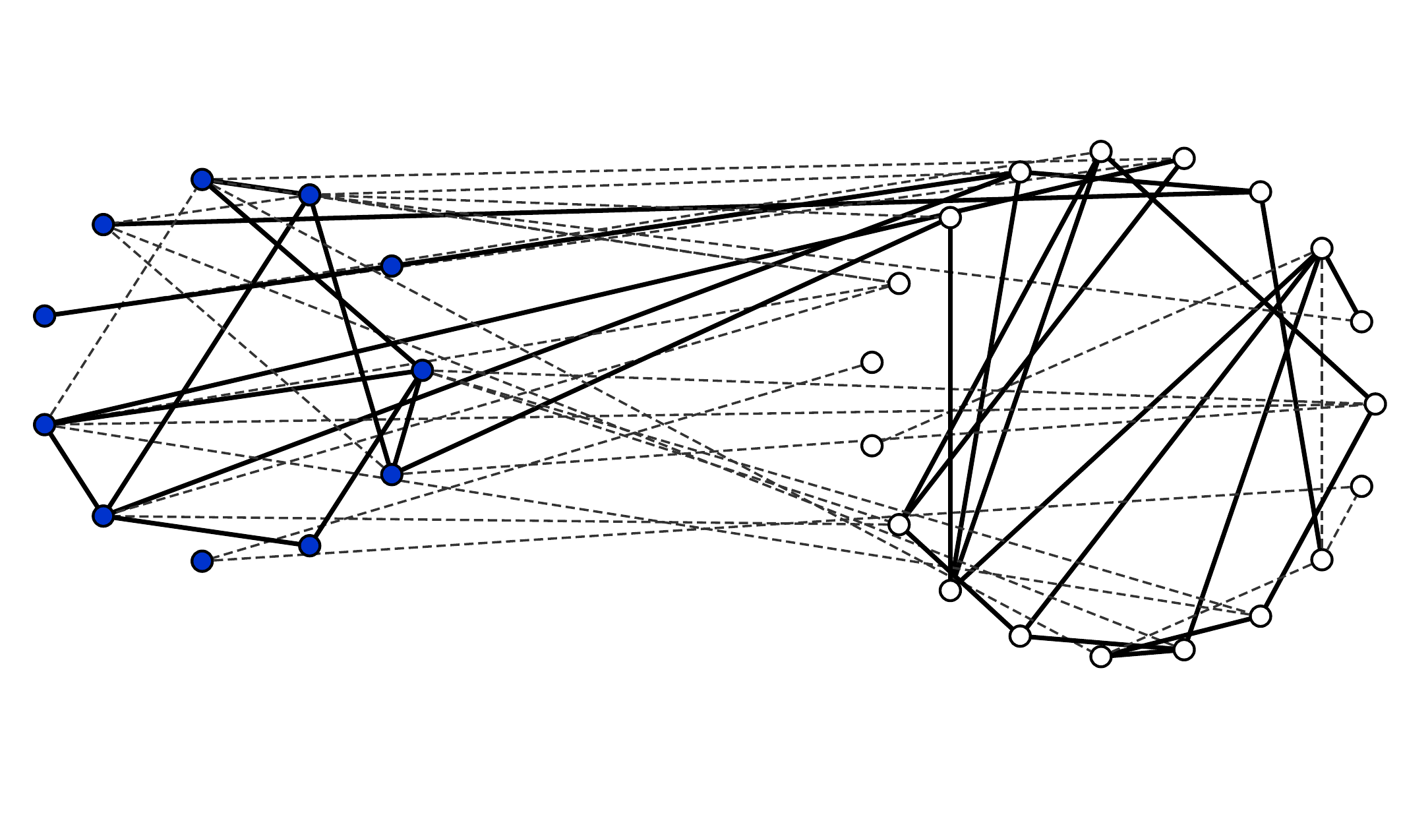}
}
\subfigure[]{
 \centering
 \includegraphics[clip, trim=0cm 15cm 0cm 0cm, width=0.48\textwidth]{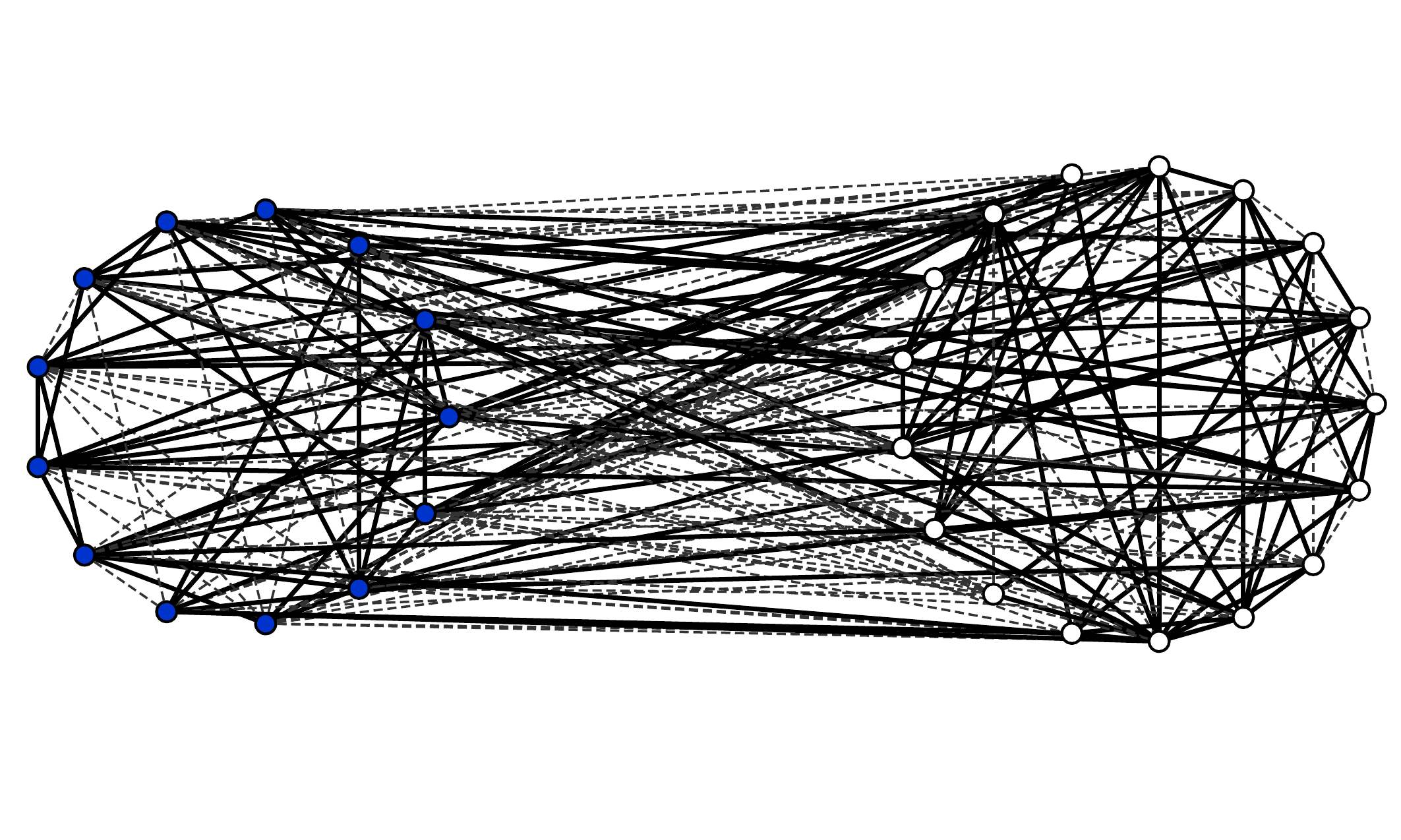}
}
 \subfigure[]{
 \centering
\includegraphics[clip, trim=0cm 15cm 3cm 0cm, width=0.48\textwidth]{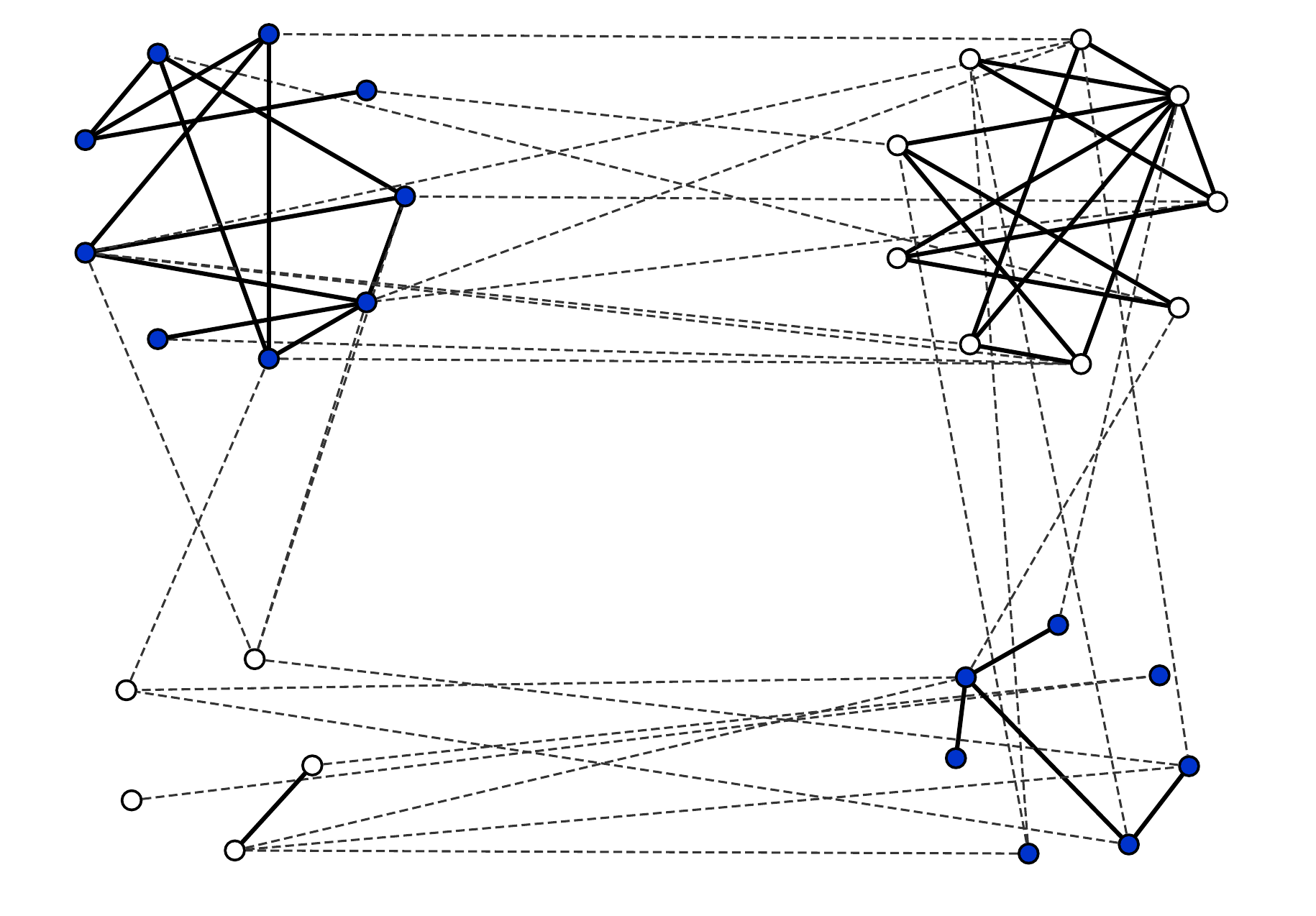}
}
\subfigure[]{
 \centering
 \includegraphics[clip, trim=0cm 15cm 0cm 0cm, width=0.48\textwidth]{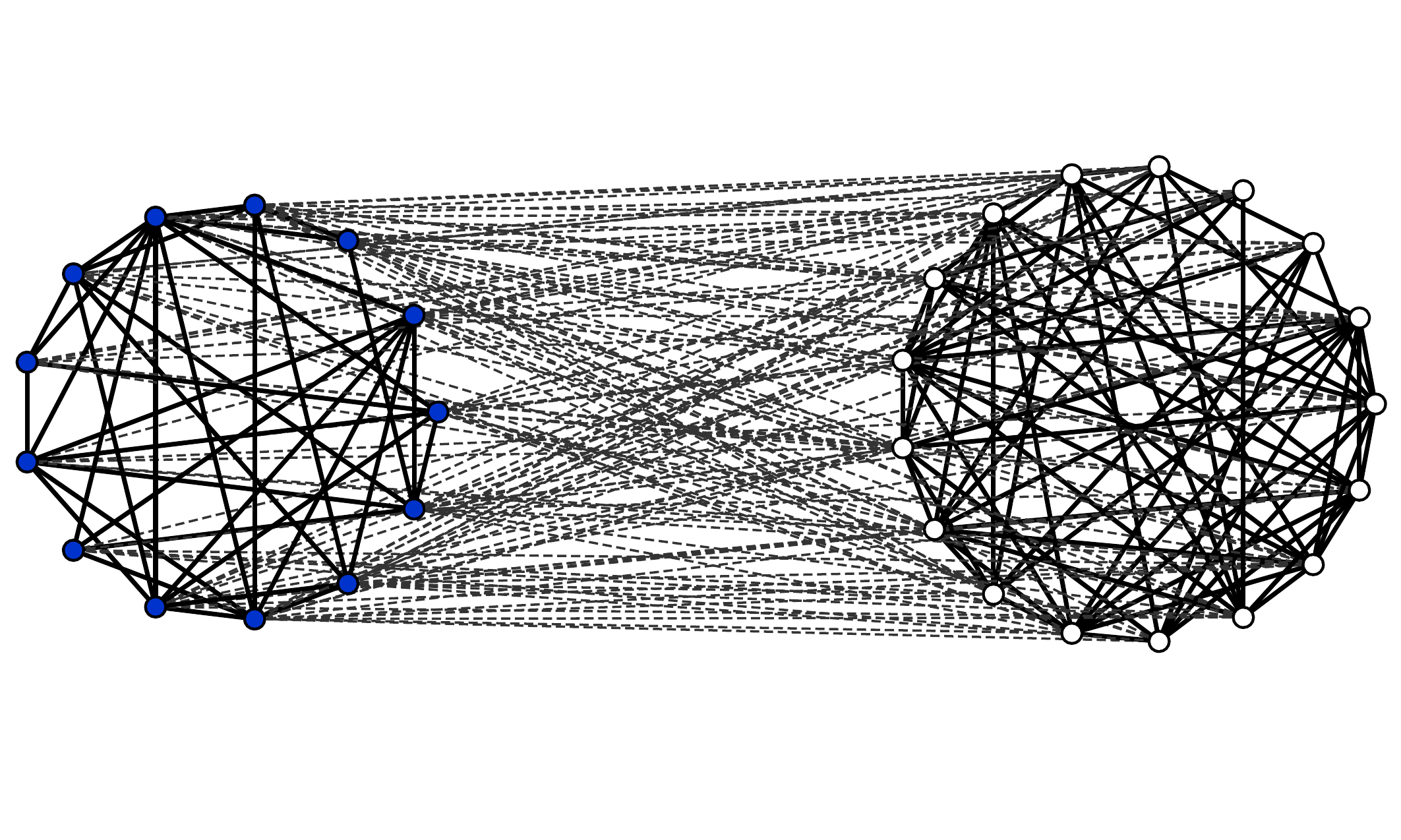}
}
\caption{In panel (a) we display a snapshot of a configuration of the dynamics with parameters $p=0.3$ and $\mu=4$ (active phase). In this configuration, all types of pair connections exist and evolve in time. This dynamics further evolves until a finite-size fluctuation takes it to the absorbing state displayed in panel (c), where there are no unsatisfying links. As it is evident from the figure, the system divides in more than two friendly groups with different opinions, a situation that corresponds to $\mu<\mu_{\rm split}\approx 14$ for these parameters. In panel (b) we show an active configuration for $p=0.7$ and $\mu=16>\mu_{\rm split}$. This configuration evolves until a fluctuation, for large times, takes it to the absorbing one displayed in panel (d). In this absorbing configuration there are only two groups, a situation corresponding to $\mu>\mu_{\rm split}$. In all cases displayed, the number of nodes is $N=30$.}
\label{group_splitting}
\end{figure}

From the topological point of view, the active phase is characterized by a continuoulsy evolving, and apparently disordered, structure of nodes a links, see Figs.~\ref{group_splitting}(a) and \ref{group_splitting}(b). The topology of the corresponding frozen phases that appear at a later time due a to a finite-size fluctuation, see Figs.~\ref{group_splitting}(c) and \ref{group_splitting}(d), will be analyzed in another section.

\section{Frozen phase}

\begin{figure}[]
\centering
 \subfigure[]{
 \centering
 \includegraphics[width=0.43\textwidth]{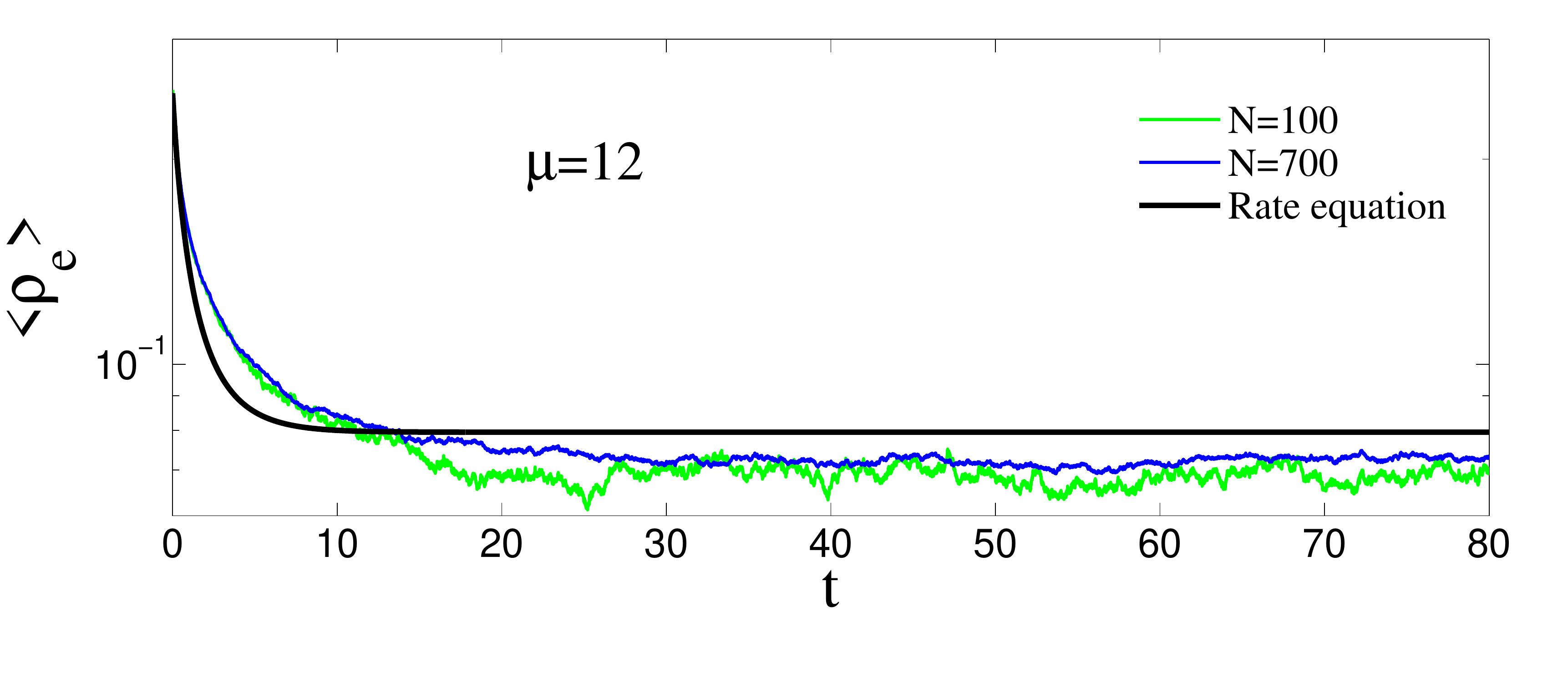}
}
 \subfigure[]{
 \centering
 \includegraphics[width=0.43\textwidth]{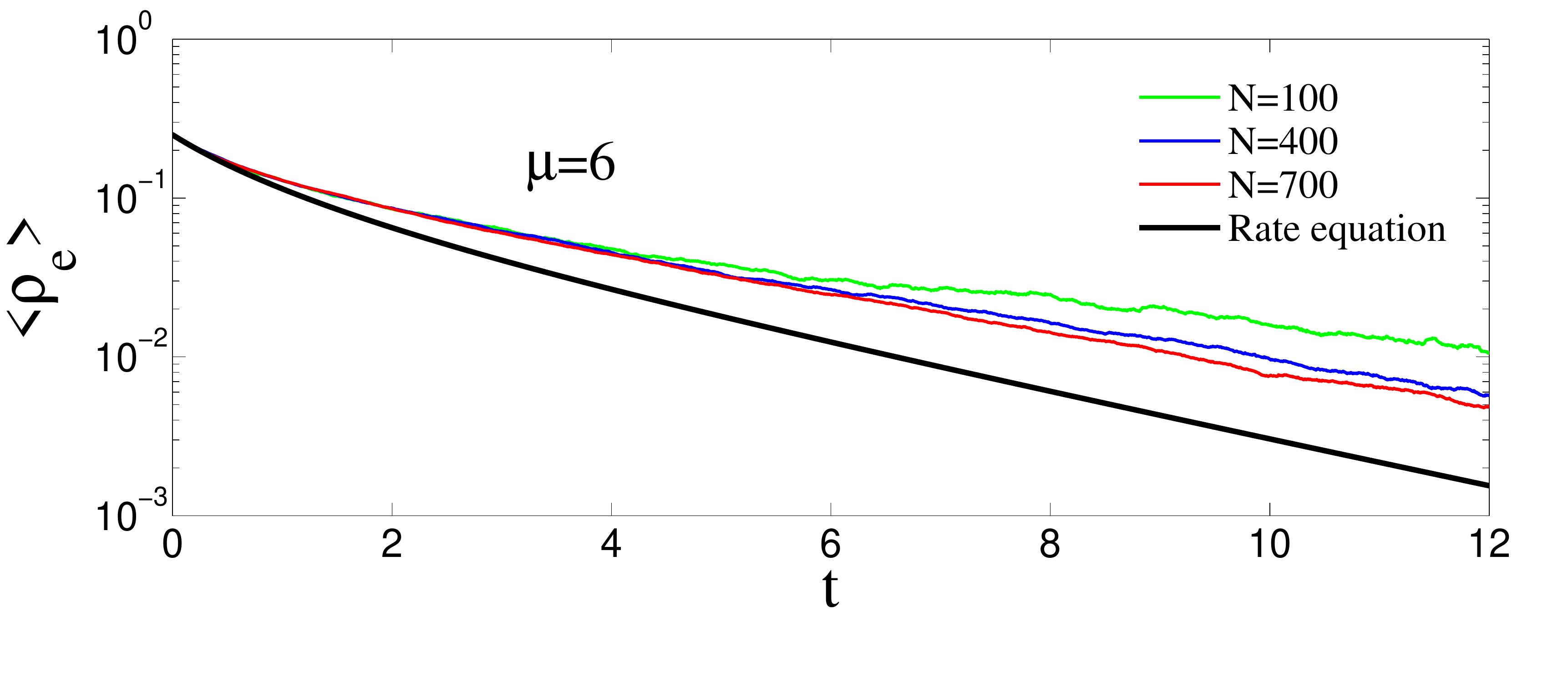}
}
\caption{Representative time evolution of $\langle\rho_{e}\rangle$ in a log-linear scale. Panel (a) corresponds to the active phase and panel (b) to the frozen phase. All trajectories in both panels use the same value of $p=0.8$ and initial condition $x_0=0.5$ and $\ell_0=0.5$ but different realizations of the network and the dynamics. In both cases, the averages use $100$ runs that did not end in the frozen phase. In panel (b) we observe an exponential behavior $\langle\rho_{e}\rangle\propto e^{-t/\tau_{0}}$ with a small dependence of $\tau_{0}$ with system size. In panel (a), the fluctuations of $\langle\rho_{e}\rangle$ around the well defined value decrease with system size.}
\label{rho_e}
\end{figure}

\begin{figure}[]
\centering
 \subfigure[]{
 \centering
 \includegraphics[width=0.235\textwidth]{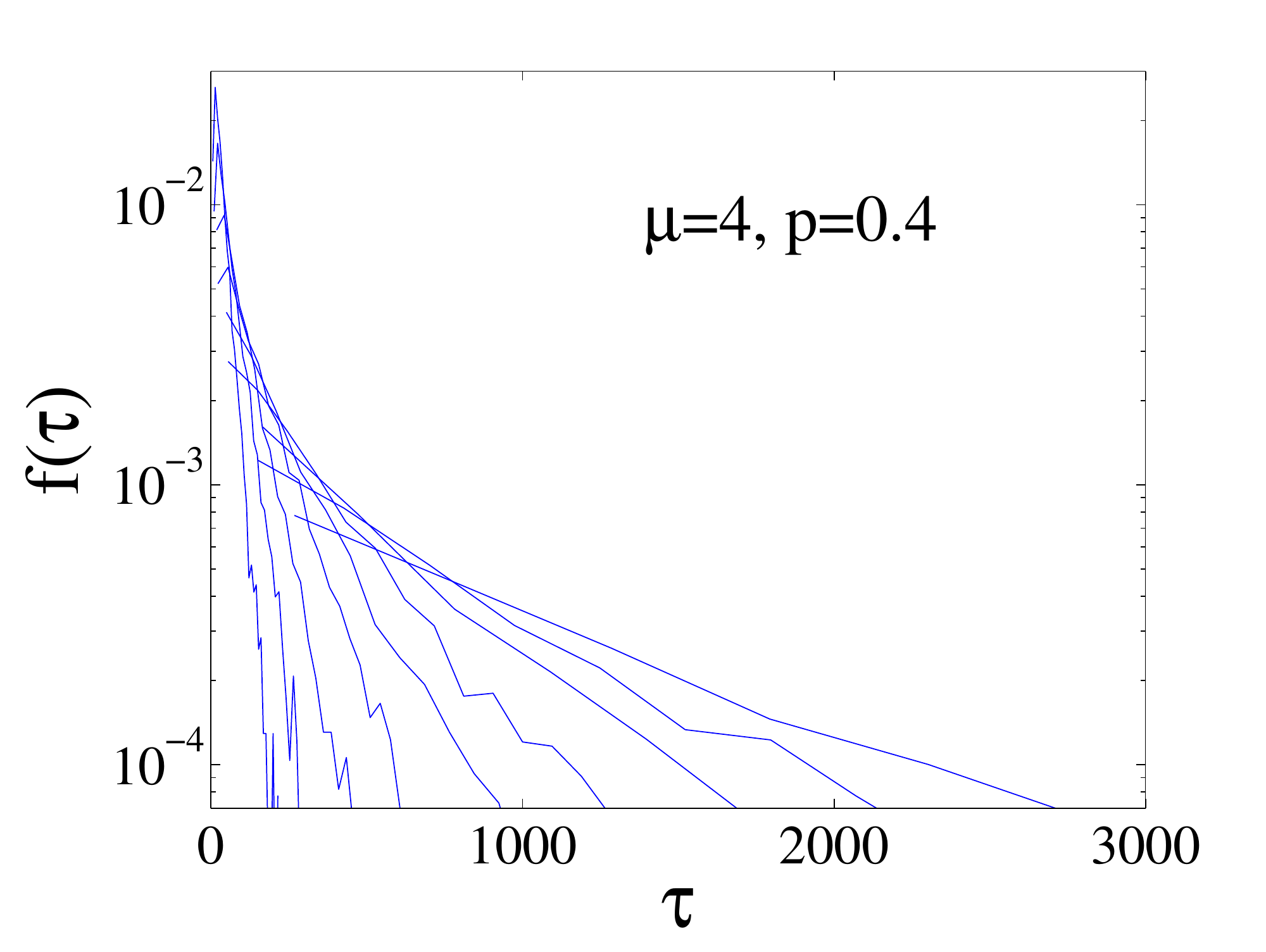}
}
 \subfigure[]{
 \centering
 \includegraphics[width=0.235\textwidth]{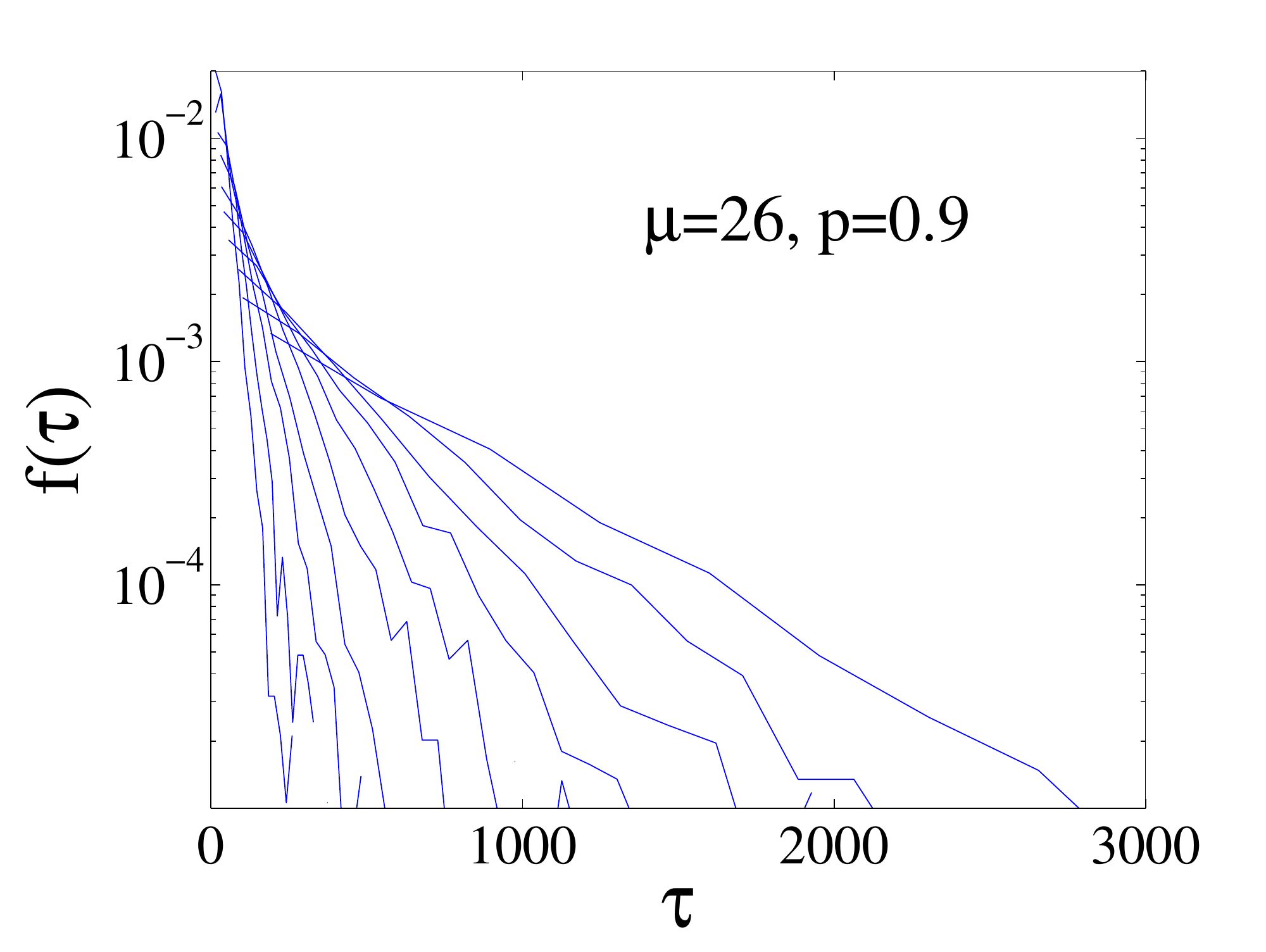}
}
 \subfigure[]{
 \centering
 \includegraphics[width=0.235\textwidth]{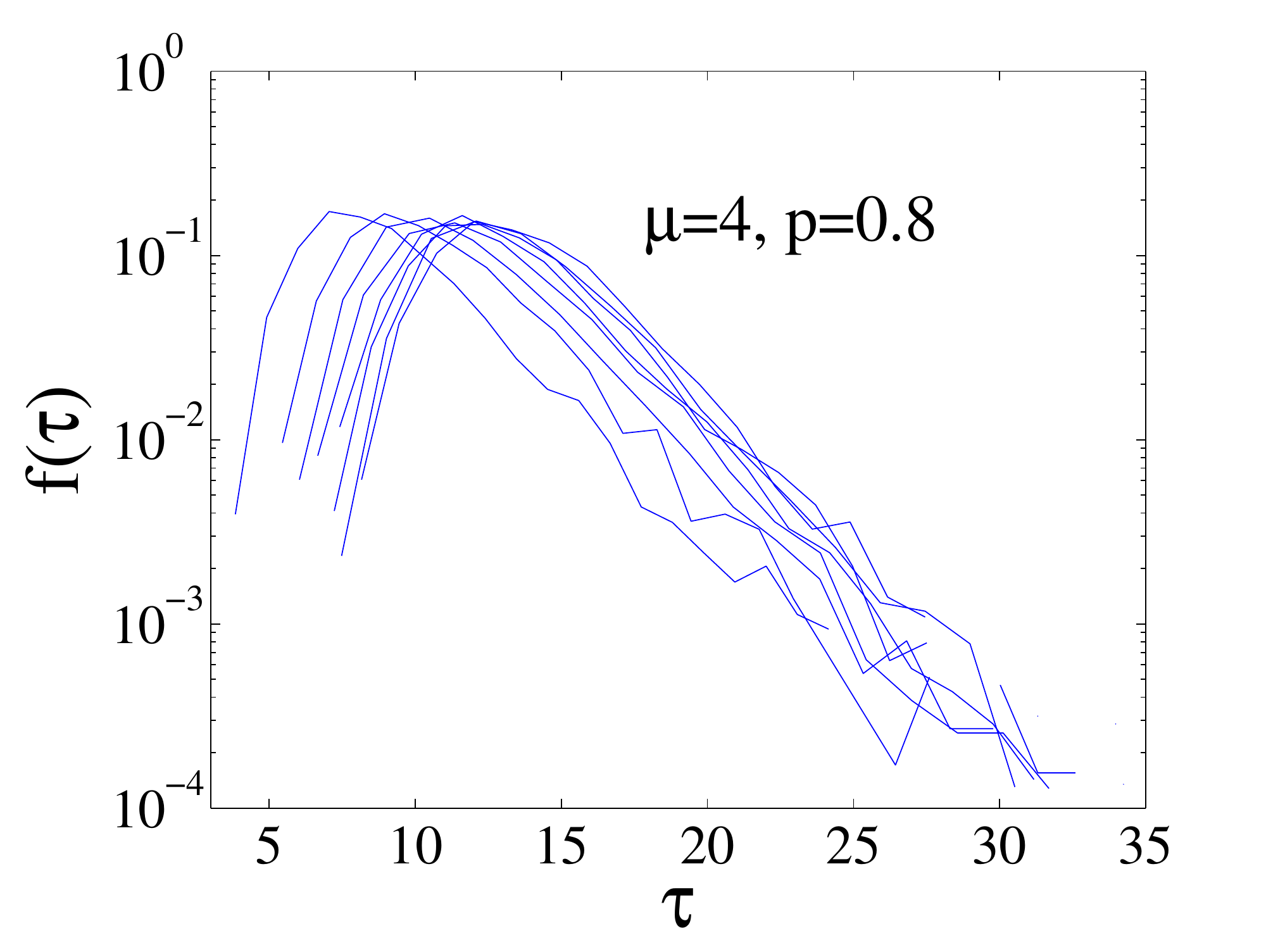}
}
 \subfigure[]{
 \centering
 \includegraphics[width=0.235\textwidth]{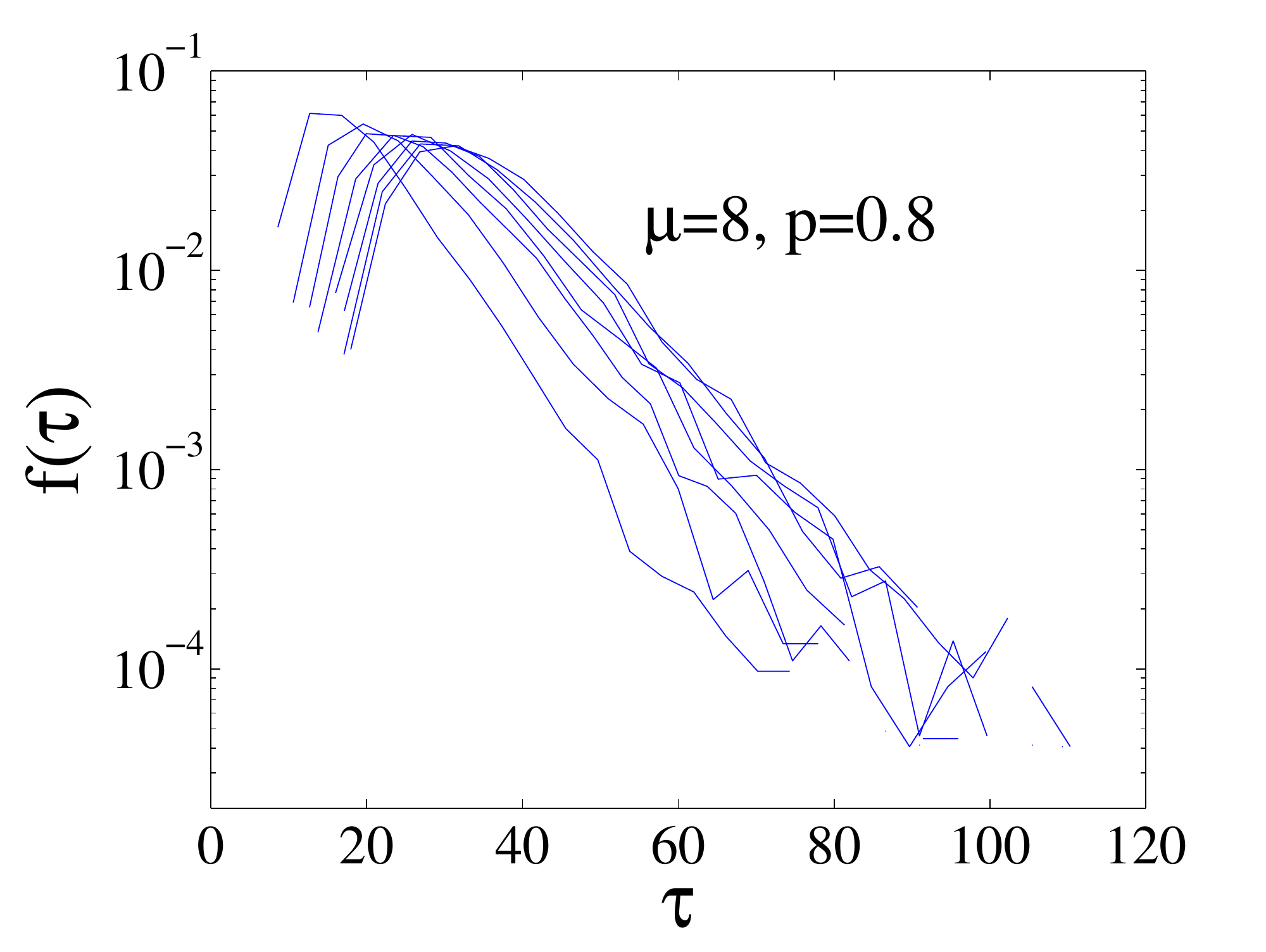}
}
 \subfigure[]{
 \centering
 \includegraphics[width=0.235\textwidth]{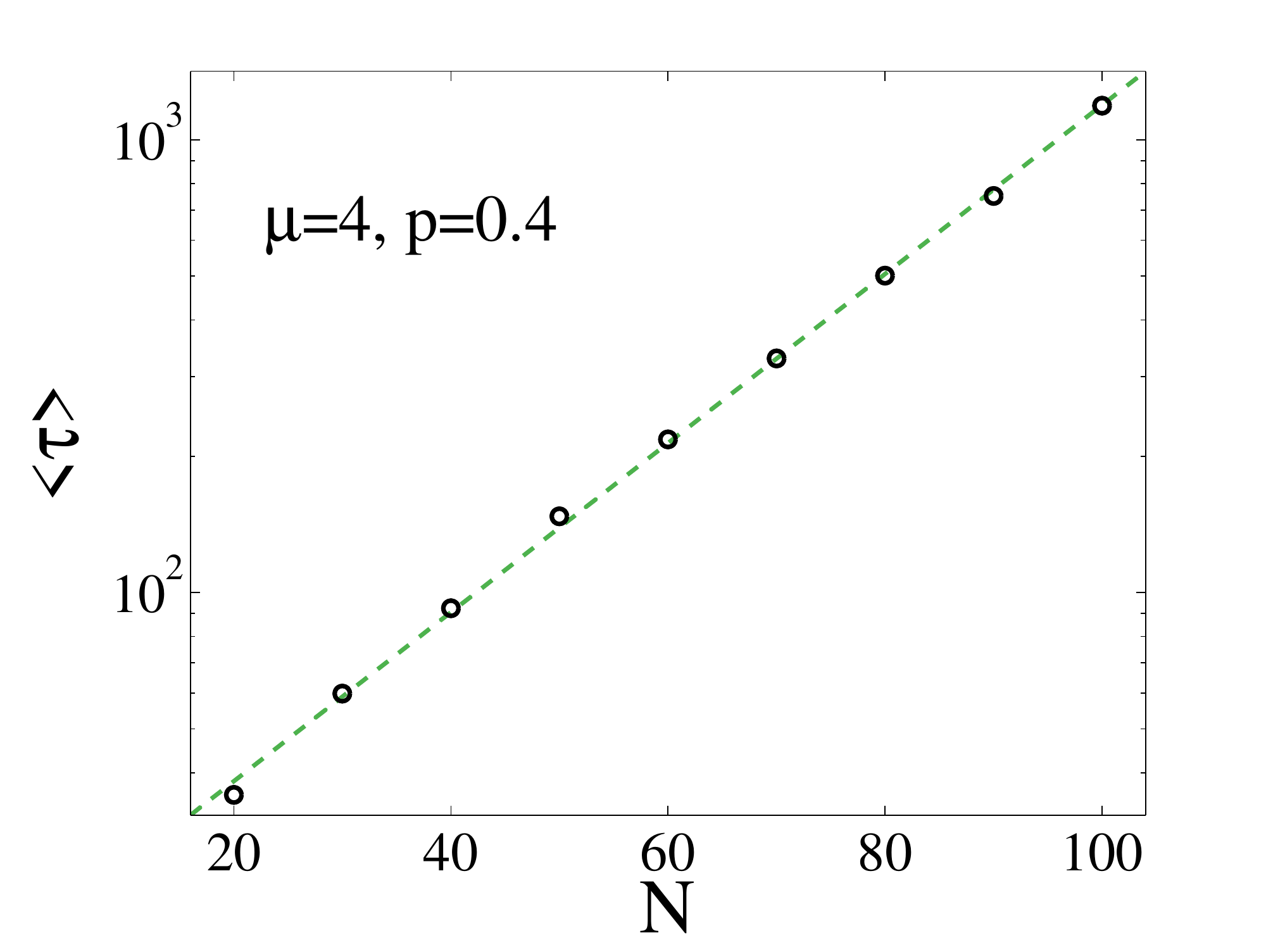}
}
 \subfigure[]{
 \centering
 \includegraphics[width=0.235\textwidth]{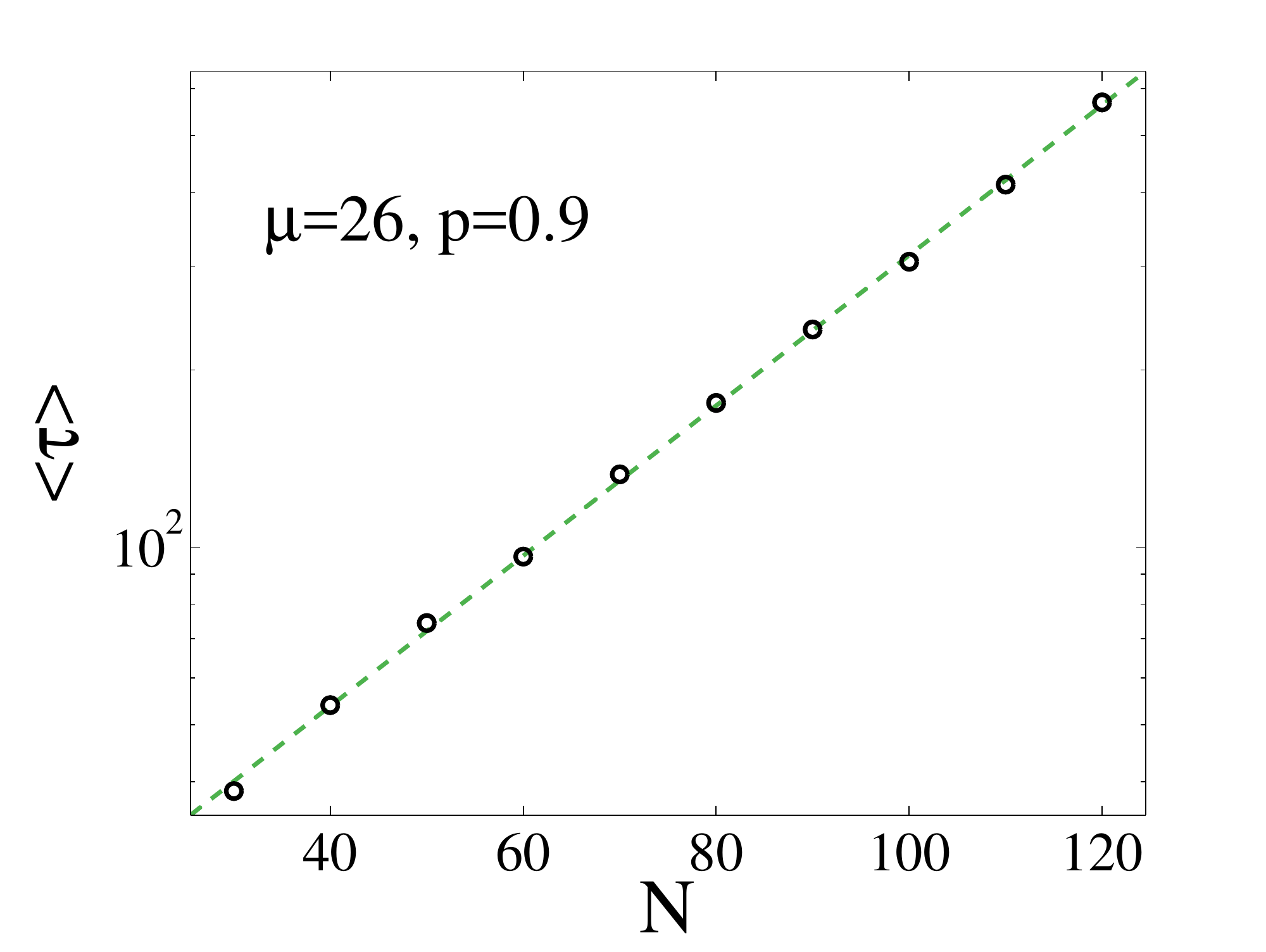}
}
 \subfigure[]{
 \centering
 \includegraphics[width=0.235\textwidth]{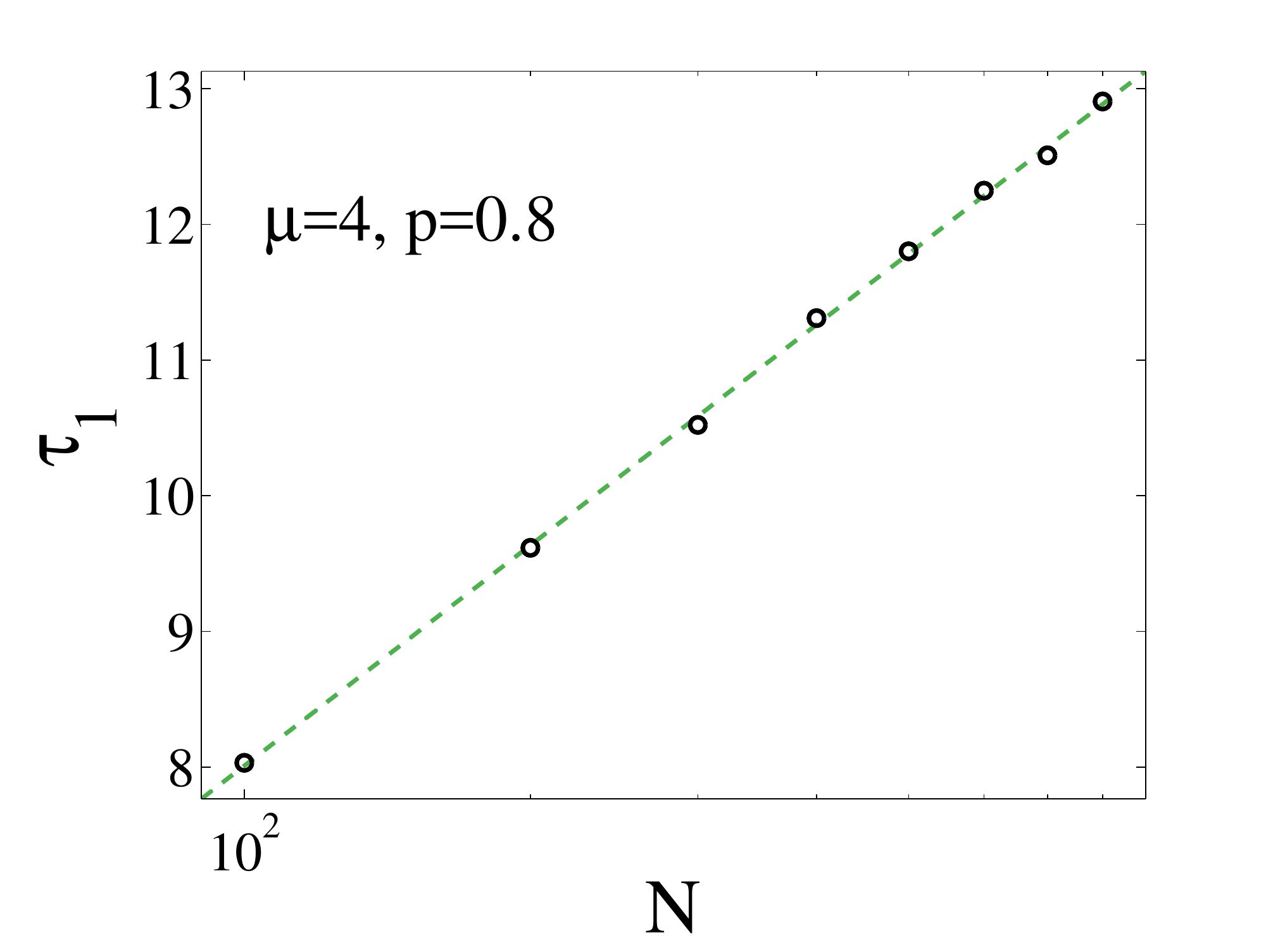}
}
 \subfigure[]{
 \centering
 \includegraphics[width=0.235\textwidth]{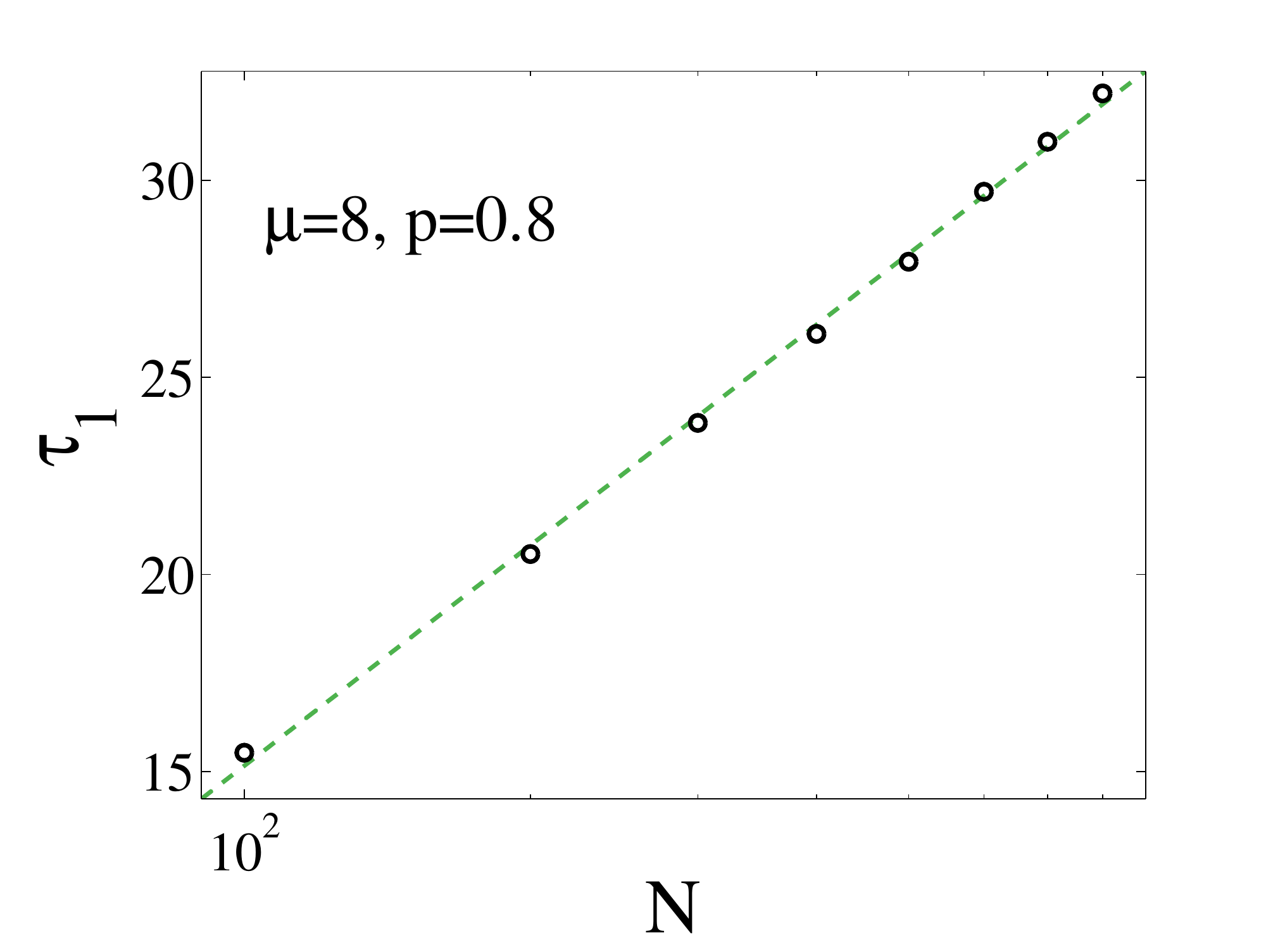}
}
\caption{In panels (a) and (b) we plot the probability density function of the characteristic time to decay to the frozen state, $f(\tau)$ for two different sets of parameters belonging to the active phase, and for different system sizes. The corresponding mean values of the distributions, $\langle\tau\rangle$ are displayed as a function of system size in panels (e) and (f), showing an exponential form, $\langle\tau\rangle\sim e^{\alpha N}$. Panels (c) and (d) plot $f(\tau)$ for two different sets of parameters belonging to the frozen phase. These distributions have well defined maximum located at $\tau_1$. The dependence of $\tau_1$ as function of system size are plotted in panels (g) and (h) displaying a logarithmic behavior, $\tau_1\sim \log(N)$.}
\label{tau}
\end{figure}

\begin{figure}[]
\centering
 \subfigure[]{
 \centering
 \includegraphics[clip, trim=0cm 15cm 0cm 0cm, width=0.49\textwidth]{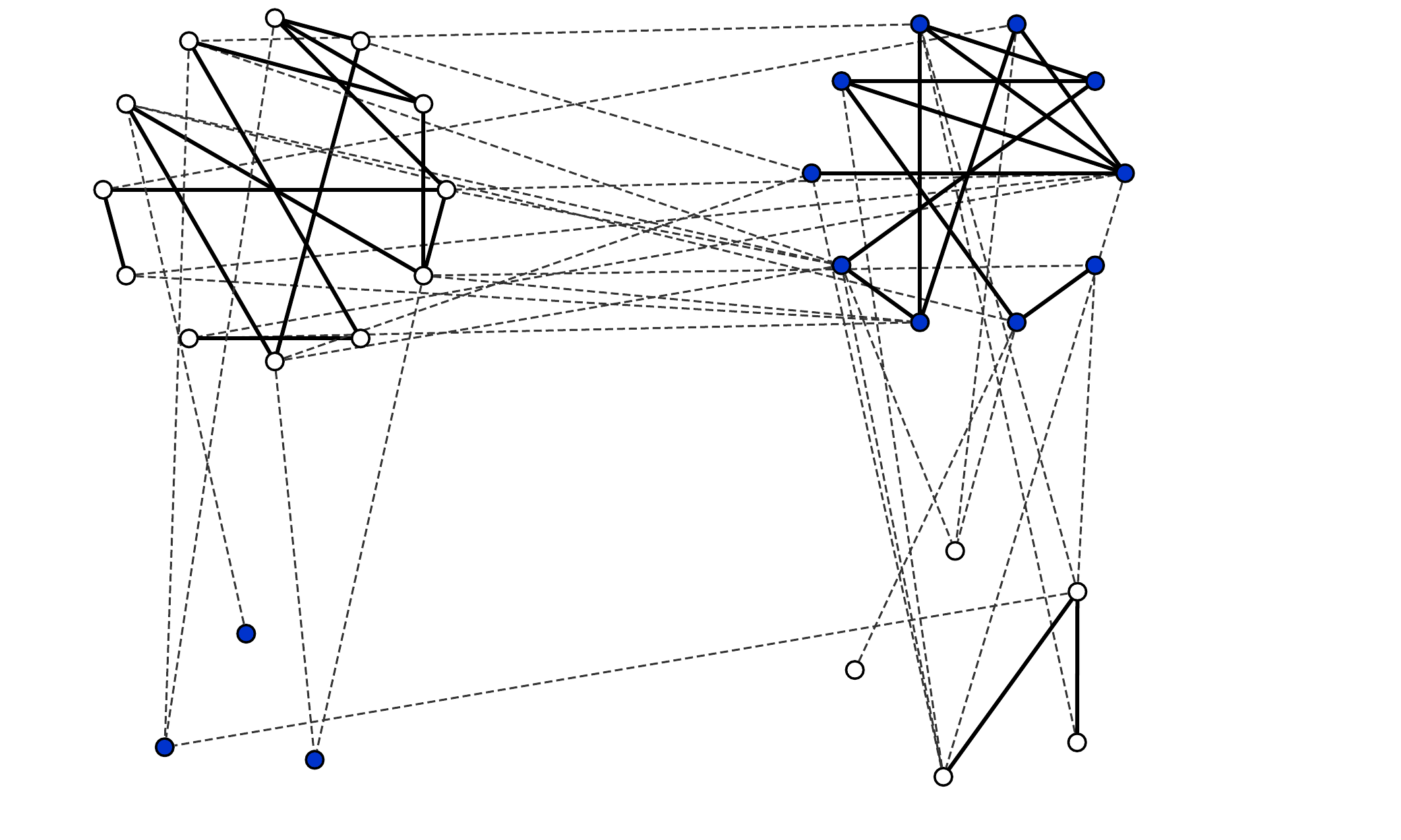}
}
 \subfigure[]{
 \centering
 \includegraphics[clip, trim=0cm 15cm 0cm 0cm, width=0.48\textwidth]{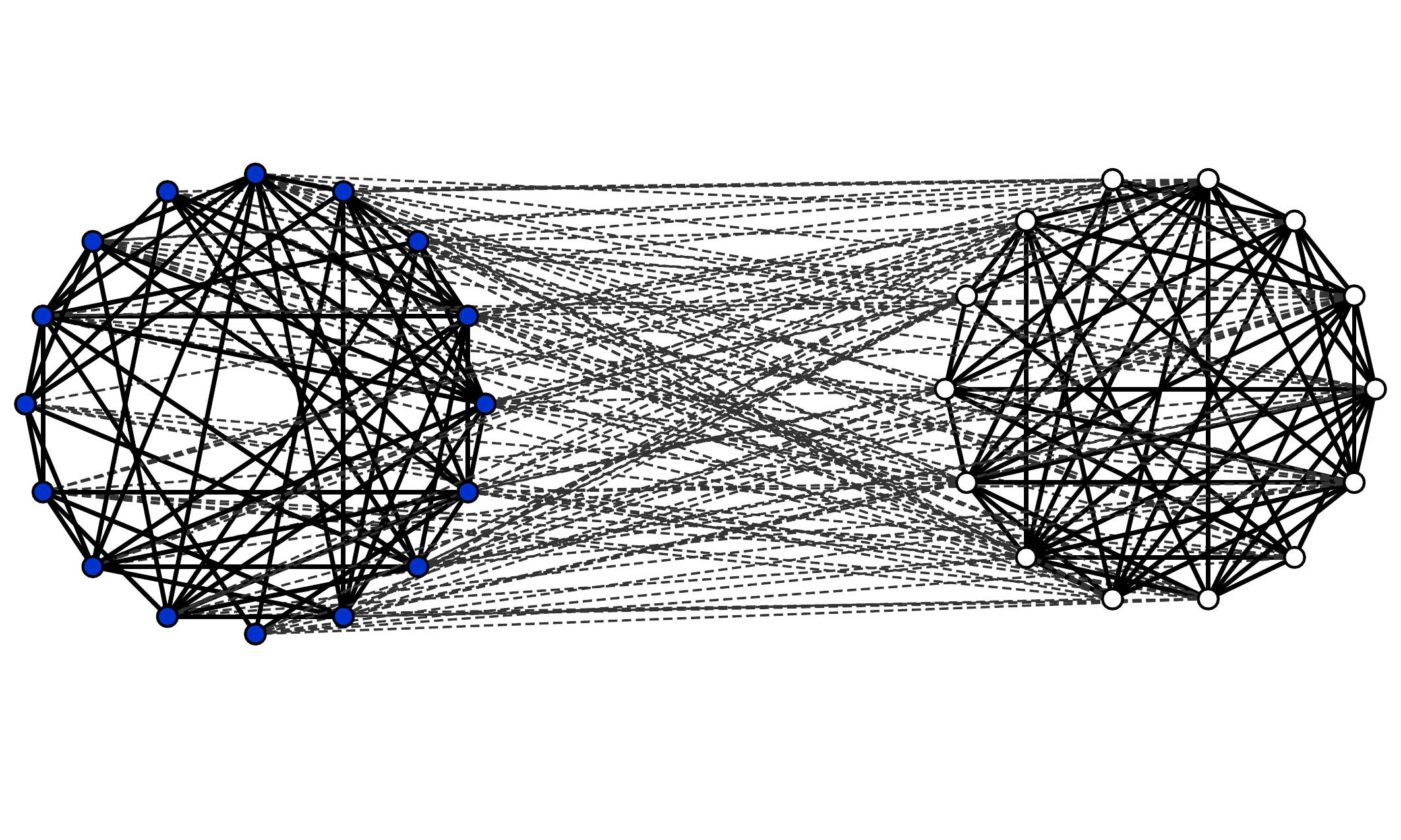}
}
\caption{In panel (a) we display a snapshot of a configuration of the dynamics with parameters $p=0.8$ and $\mu=4$ that has evolved into the frozen state with more than two groups, a situation corresponding to $\mu<\mu_{\rm split}$. Panel (b), with parameters $p=0.95$, $\mu=16$, has evolved into a frozen phase with two groups, as it corresponds to $\mu>\mu_{\rm split}$. In both cases, the number of nodes is $N=30$.}
\label{group_splitting-f}
\end{figure}

When the link-update probability is larger than the critical value $p>p_c(\mu)$ the system falls into the absorbing or frozen phase. At variance with the system in the active phase, the densities of unsatisfied links $\rho_a(t)$, $\rho_c(t)$ and $\rho_e(t)$ continuously decrease during the time evolution and never reach a plateau from which they eventually escape. Therefore, there is a continuous decay towards the absorbing phase, contrarily to the decay of the active phase that was produced by a rare fluctuation. This is evidenced in panel (b) of Fig.~\ref{rho_e} where we plot the time evolution of $\rho_e(t)$ for different system sizes. We observe an exponential decay $\rho_e(t)\sim e^{-t/\tau_0}$ with a very small dependence of $\tau_0$ on the system size $N$ and approaching a limiting value relatively close to the prediction of the rate equation. The time to reach the absorbing state, $\tau$, is also a random variable characterized by a pdf $f(\tau)$. As shown in panels (c) and (d) Fig.~\ref{tau}, corresponding to two different points in the frozen phase, the tail of $f(\tau)$ can still be fitted by an exponential function $e^{-t/\tau_2}$, but this function now presents a well defined maximum located at $\tau_1$, the characteristic time for decay unto the frozen state. This characteristic time is plotted in panels (g) and (h) of Fig.~\ref{tau} showing a logarithmic increase with system size $\tau_1\sim\log N$. It is possible to relate the exponential decay observed in $\rho_e(t)$ with this logarithmic dependence. The transition to the absorbing state will occur at the time $\tau_1$ when the density of unsatisfying pairs falls below a value of order $1/N$, i.e. $\rho_e(\tau_1)\sim 1/N$. Replacing $\rho_e(t)\sim e^{-t/\tau_0}$. We arrive at $\tau_1\sim\tau_0\log N$, a logarithmic dependence on system size, as observed. Snapshot of typical frozen configurations are displayed in Figs.~\ref{group_splitting-f}(a) and \ref{group_splitting-f}(b). In the next section we discuss the possible topological structures of the frozen phases.

\section{Finite size topological transition}

A fully satisfying, absorbing, configuration obtained in an Erd\"os-R\'enyi network displays a transition associated with some structure that can be described as ``group splitting''. By that, we mean that the nodes organize in several groups, defining a group as a set of nodes holding the same opinion and connected by friendly links among themselves and by unfriendly links to the members of other groups. This group splitting structure appears both when the absorbing configuration has been reached from a finite-size fluctuation of an active phase, $p<p_c(\mu)$, see Figs.~\ref{group_splitting}(c) and \ref{group_splitting}(d), or when it corresponds to the frozen phase in parameter space, $p>p_c(\mu)$, see Figs.~\ref{group_splitting-f}(a) and \ref{group_splitting-f}(b).

\begin{figure}[]
\centering
 \subfigure[]{
 \centering
 \includegraphics[width=0.31\textwidth]{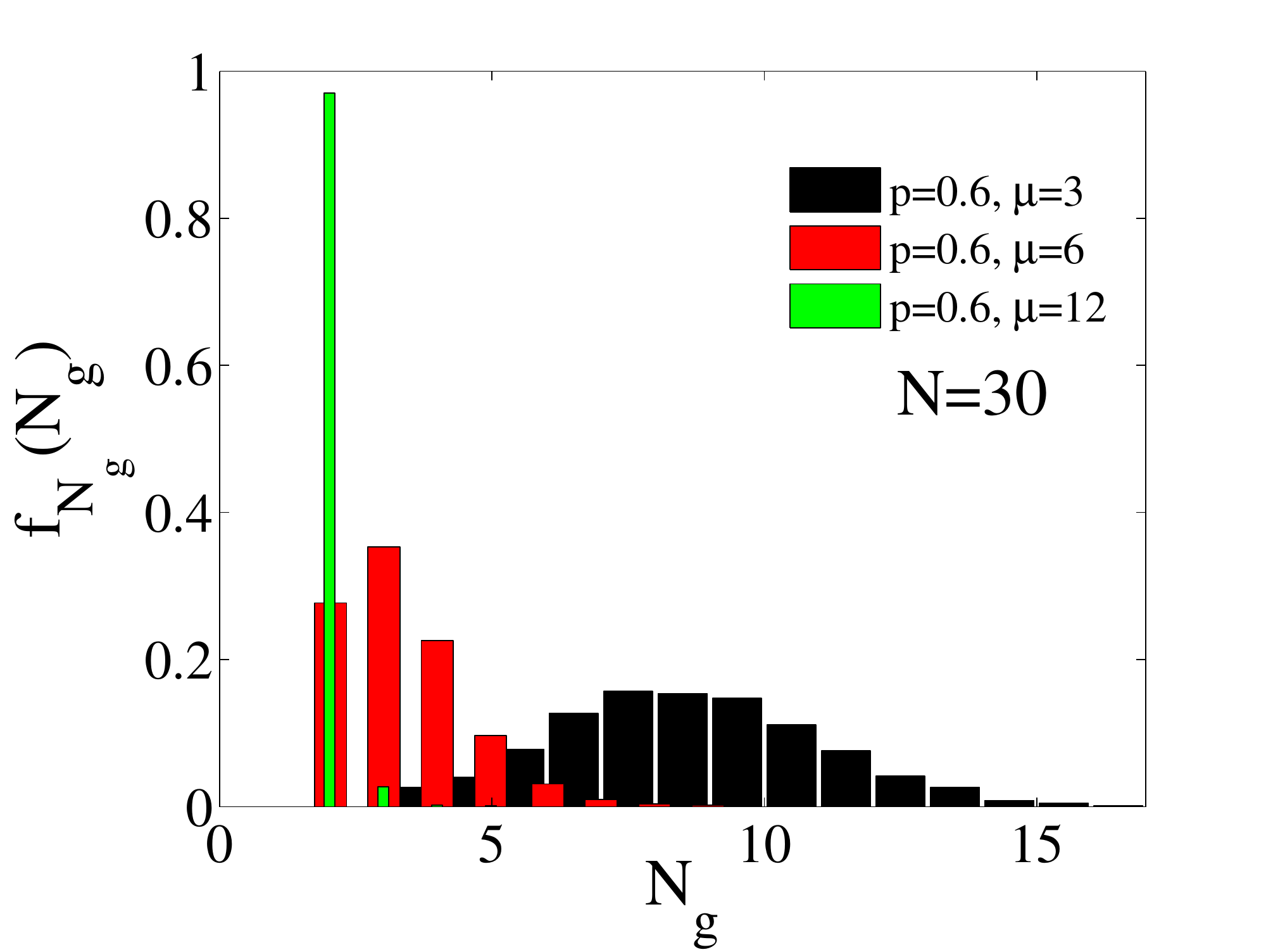}
}
 \subfigure[]{
 \centering
 \includegraphics[width=0.31\textwidth]{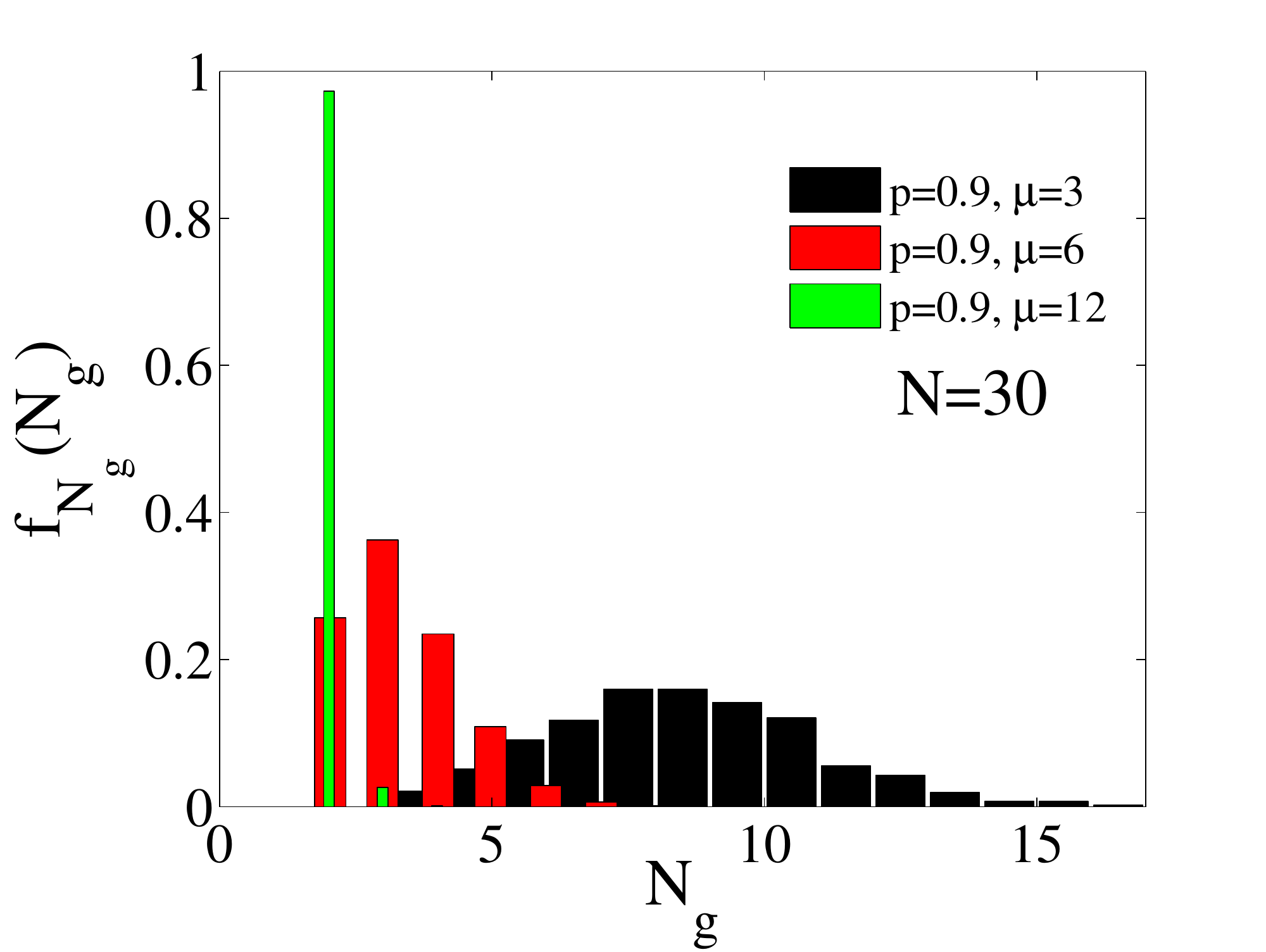}
}
 \subfigure[]{
 \centering
 \includegraphics[width=0.31\textwidth]{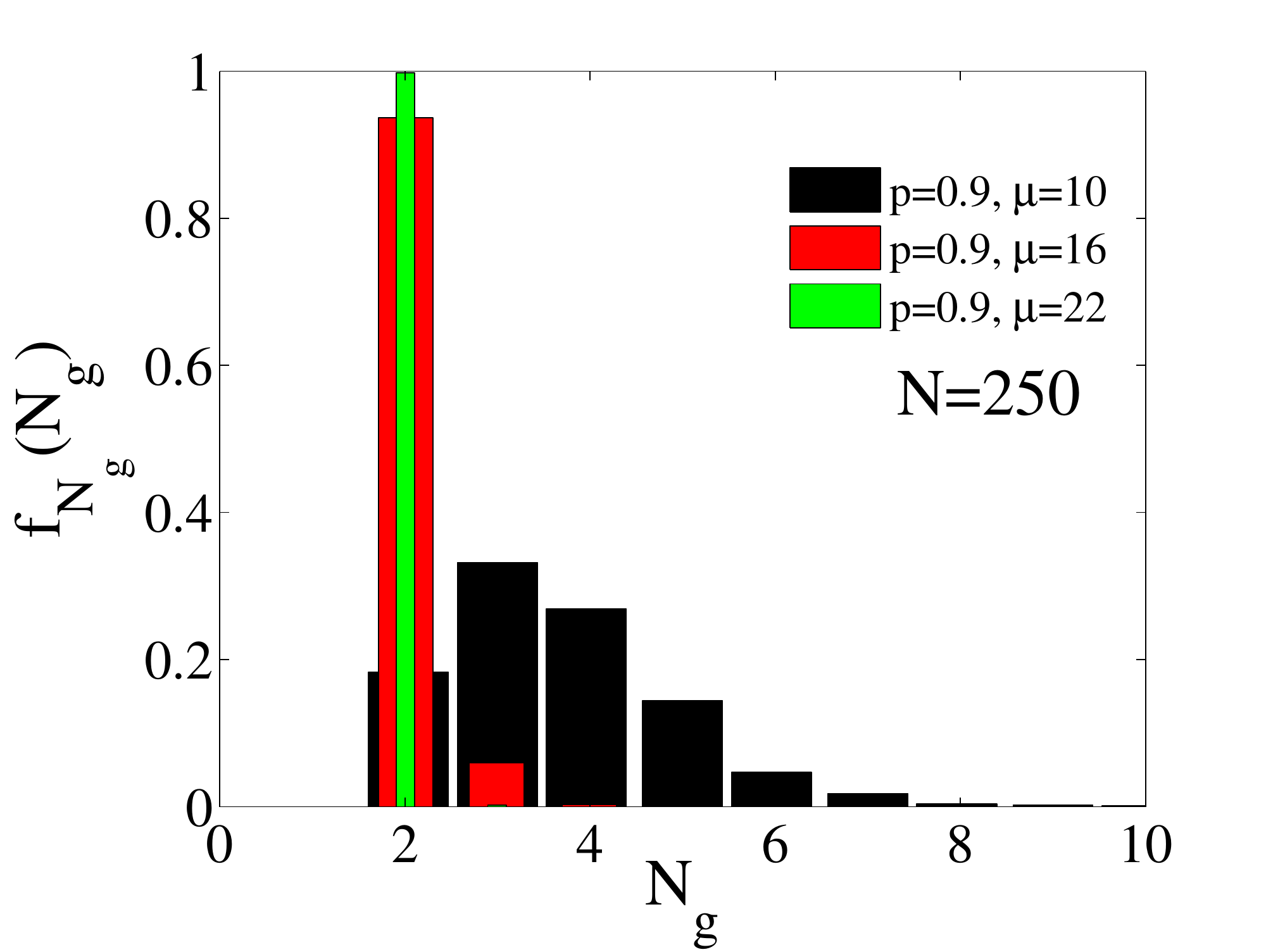}
}
 \caption{Probability density function of the number of groups for fixed $p$ but different $\mu$ in panels are presented. The distributions with the same color in panel (a) and (b) correspond to the same $\mu$ but different $p$ for $x_0=0.5$, $\ell_0=0.5$ and as can be seen, roughly speaking, they are indistinguishable which means that the mean value and variance of $f_{N_g}$ don't depend on $p$ but on $\mu$, in mentioned initial condition. If we change the initial condition those also will depend on $p$ which is not our interest. In addition we can see that in panel (a) and (b) with $N=30$ the critical value for splitting is $\mu<\mu_{\rm split}\backsimeq14$ but in the panel (c) with $N=250$ critical value is $\mu<\mu_{\rm split}\backsimeq22$. That means the critical value increase with system size. }
\label{pdf_N}
\end{figure}

We characterize the absorbing configurations by the number of groups $N_g$. This number is a stochastic variable that depends on the initial condition, the system parameters and the particular realization of the dynamics. We denote by $f_{N_g}(N_g)$ the probability distribution function of this variable. What we have observed is that there is a topological transition in which $f_{N_g}(N_g)$ changes from a unimodal distribution at $N_g=2$ (meaning that all realizations end up in two groups) to a wide distribution in which different realizations reach a state in which two large groups coexist with a varying number of smaller groups. Examples of two groups-splitting can be seen in Fig.\ref{group_splitting}(d) and Fig.\ref{group_splitting-f}(b), whereas more than two groups are displayed in Fig.\ref{group_splitting}(c) and Fig.\ref{group_splitting-f}(a). The topological transition appears when crossing the line $\mu_{\rm split}(p)$ in parameter space $(\mu,p)$ such that for $\mu>\mu_{\rm split}$ the system always fall into exactly two friendly group with different opinion, $N_g=2$, but for $\mu<\mu_{\rm split}$, the system splits in more than two friendly groups, $N_g>2$. The exact location of the transition line $\mu_{\rm split}(p)$ depends on the initial condition and on system size $N$. The simulations indicate that for the initial condition $(x_0=0.5$, $\ell_0=0.5)$ the value $\mu_{\rm split}$ is roughly independent on $p$, see panels (a) and (b) in Fig.~\ref{pdf_N}. Furthermore the transition point $\mu_{\rm split}$ grows with system size, see panel (c) of Fig.~\ref{pdf_N}, and we speculate that it tends to infinity with $N$, in such a way that in the thermodynamic limit, a typical absorbing configuration always contains more than two groups. For a different set of initial conditions, it is not true that the line $\mu_{\rm split}(p)$ is independent of $p$, but the same conclusion is reached about the disappearance of the two-groups phase in the large $N$ limit.

\begin{figure}[]
\centering
\includegraphics[width=0.43\textwidth]{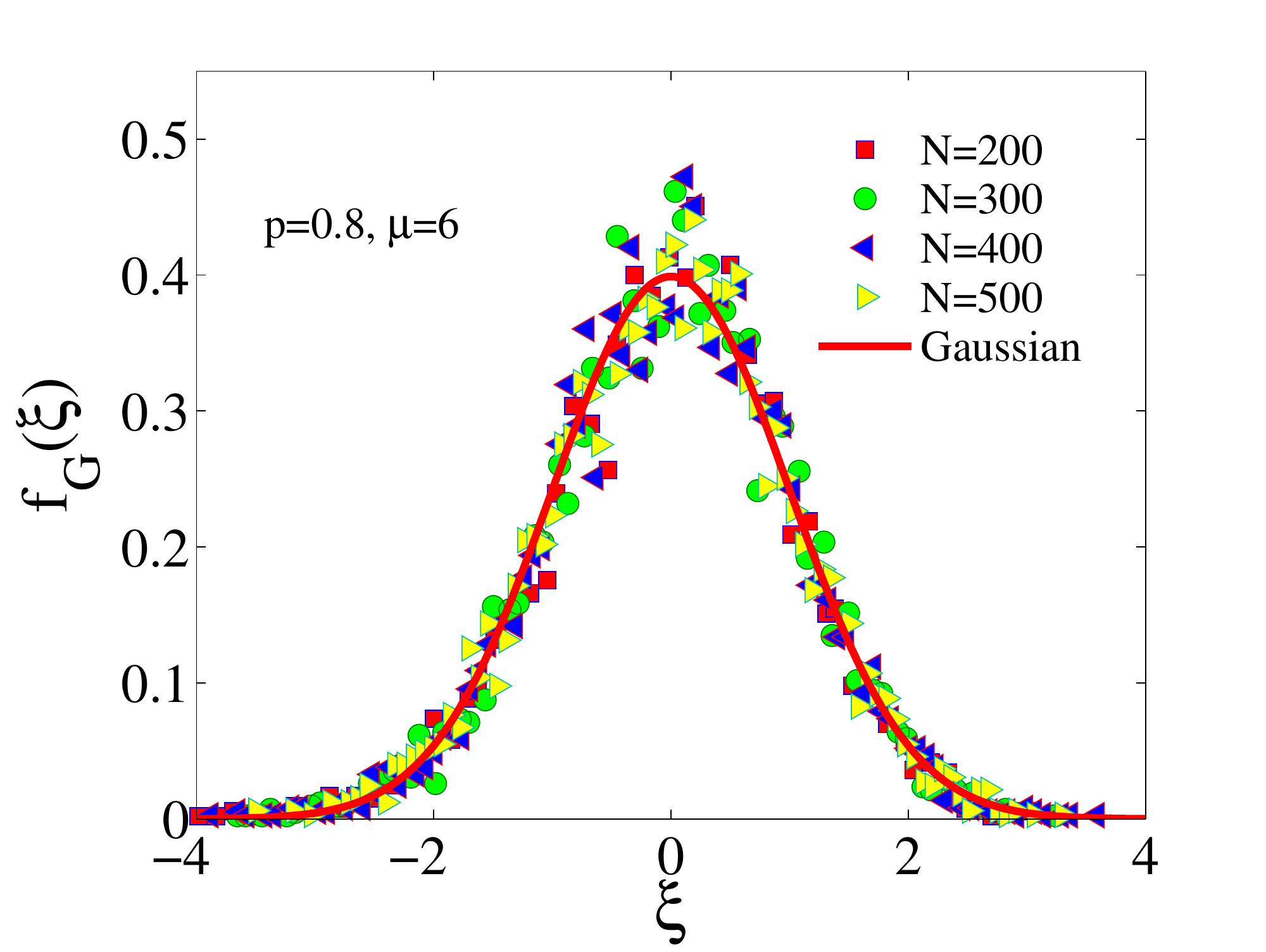}
\caption{Probability density function of the size of the largest group of the frozen phase for $p=0.8$ and $\mu=6$. The results for different system sizes have been rescaled by defining $\xi=(G-\langle G\rangle)/\sigma[G]$, that can be clearly fitted by a Gaussian distribution (solid line). In all cases the initial condition is $x_0=0.5$, $\ell_0=0.5$.}
\label{pdf_G}
\end{figure}

The size $G$ of the largest white group is also a random variable described by the corresponding pdf $f_G(G)$. The mean value and variance, $\langle G\rangle$, $\sigma^2[G]$ of that distribution depend, besides $p$ and $\mu$, on the initial densities of white opinions $x_0$ and friendly links $\ell_0$, in a similar functional form that the final density $\langle x\rangle^\text{st}$ displayed in Fiq.~\ref{lattice2}(c) and (d). For a given initial condition, $\langle G\rangle$ and $\sigma^2[G]$ increase linearly with system size $N$ (not shown). In Fig.~\ref{pdf_G} we show that $f_G(G)$ can be well represented by a Gaussian distribution.

\section{Summary and Discussion}
 We have introduced a model of opinion formation in the context of the study of coupled dynamics of node and link states in a complex network: We postulate that friendly/unfriendly links can affect the process of changing opinions so that friends like to have the same opinion and unfriendly relations are satisfied with different opinions. We have proposed a dynamical rule for the evolution of unsatisfied pairwise relations to satisfactory relations by either node or link updates. The relevant parameter of the problem $p$ is the probability for link update instead of a node update in the local dynamics rule. By a mean-field rate equation analysis, corroborated by Monte Carlo simulations, we find an absorbing continuous phase transition from a frozen to a dynamically active phase occurring for a critical value of $p$. In spite of a dynamical rule of local convergence, global convergence to the satisfactory absorbing phase does not occur for slow link update. In the active phase, the density of all possible pairwise relations fluctuate around well defined values that depend on $\mu$ and $p$. However, finite-size fluctuations take the system to a frozen configuration, but this occurs in a characteristic time that grows exponentially with system size. In the frozen phase, the system orders dynamically, with an order parameter decaying exponentially to zero. For a finite system, the characteristic time to reach the absorbing state in this phase grows logarithmically with system size. The final frozen configuration reached by finite-size fluctuations either in the active or absorbing phases shows a group structure such that the links within a group are friendly and the links between groups are unfriendly. There is a finite-size topological transition between a two group and a multigroup structure of those final frozen configurations.

\appendix

\section{Method}
\subsection{Rate equation of coupled evolution of node and link in imitating process in the mean-field approximation}
To predict the behavior of the different densities of pairs as a function of time, we derive the rate equations of the dynamic sketched in Fig.~\ref{update_rule} on a network in which each nodes has exactly $\mu$ links as
\begin{eqnarray}
\frac{d\rho_{a}}{dt}&=&-\rho_{a}+(1-p)(\mu-1)\Big(-\frac{\rho_{a}^{2}}{\chi}+\frac{1}{2}\frac{\rho_{c}\rho_{f}}{\phi}+\frac{1}{2}\frac{\rho_{e}\rho_{f}}{2\phi}-\frac{\rho_{e}\rho_{a}}{2\chi}\Big) \label{eq:rate} \\
\frac{d\rho_{b}}{dt}&=& p\rho_{a}+\frac{(1-p)}{2}\rho_{e}+(1-p)(\mu-1)\Big(-\frac{\rho_{a}\rho_{b}}{\chi}+\frac{1}{2}\frac{\rho_{c}\rho_{e}}{\phi}+\frac{1}{2}\frac{\rho_{e}^{2}}{2\phi}-\frac{\rho_{e}\rho_{b}}{2\chi}\Big) \nonumber\\
\frac{d\rho_{c}}{dt}&=&-\rho_{c}+(1-p)(\mu-1)\Big(\frac{1}{2}\frac{\rho_{a}\rho_{f}}{\chi}-\frac{\rho_{c}^{2}}{\phi}+\frac{1}{2}\frac{\rho_{e}\rho_{f}}{2\chi}-\frac{\rho_{e}\rho_{c}}{2\phi}\Big) \nonumber \\
\frac{d\rho_{d}}{dt}&=& p\rho_{c}+\frac{(1-p)}{2}\rho_{e}+(1-p)(\mu-1)\Big(\frac{1}{2}\frac{\rho_{a}\rho_{e}}{\chi}-\frac{\rho_{c}\rho_{d}}{\phi}+\frac{1}{2}\frac{\rho_{e}^{2}}{2\chi}-\frac{\rho_{e}\rho_{d}}{2\phi}\Big) \nonumber\\
\frac{d\rho_{e}}{dt}&=& -\rho_{e}+(1-p)(\mu-1)\Big(-\frac{1}{2}\frac{\rho_{a}\rho_{e}}{\chi}+\frac{\rho_{a}\rho_{b}}{\chi}+\frac{\rho_{c}\rho_{d}}{\phi}-\frac{1}{2}\frac{\rho_{c}\rho_{e}}{\phi}+\frac{\rho_{e}\rho_{d}}{2\phi}-\frac{1}{2}\frac{\rho_{e}^{2}}{2\chi}-\frac{1}{2}\frac{\rho_{e}^{2}}{2\phi}+\frac{\rho_{e}\rho_{b}}{2\chi}\Big) \nonumber \\
\frac{d\rho_{f}}{dt}&=&p\rho_{e}+(1-p)(\rho_{a}+\rho_{c})\nonumber\\&&+(1-p)(\mu-1)\Big(-\frac{1}{2}\frac{\rho_{a}\rho_{f}}{\chi}+\frac{\rho_{a}^{2}}{\chi}+\frac{\rho_{c}^{2}}{\phi}-\frac{1}{2}\frac{\rho_{c}\rho_{f}}{\phi}+\frac{\rho_{e}\rho_{c}}{2\phi}-\frac{1}{2}\frac{\rho_{e}\rho_{f}}{2\chi}-\frac{1}{2}\frac{\rho_{e}\rho_{f}}{2\phi}+\frac{\rho_{e}\rho_{a}}{2\chi} \Big)\nonumber \\
\chi&=&\rho_{a}+\rho_{b}+\rho_{e}+\rho_{f} \nonumber \\
\phi&=&\rho_{c}+\rho_{d}+\rho_{e}+\rho_{f} \nonumber
\end{eqnarray}
According to the update rule, the only active pairs are $a$, $c$ and $e$, however due to the node update the statues of pairs $b$, $d$ and $f$ will change. Because, in the procedure of node update, the states of all pairs connected to the updated node will also be changed. The nonlinear terms in the rate equation are the consequence of this interaction. In the following we explain the derivation of the rate equation.

\begin{figure}\centering
\includegraphics[width=0.6\linewidth]{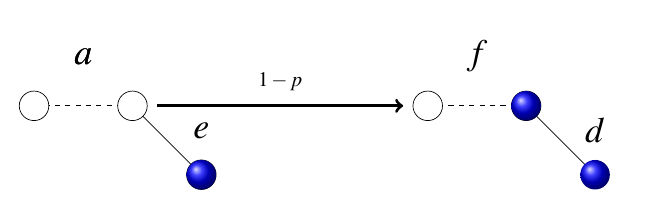}
\caption{The pair \textit{e} becomes \textit{d} when a node update from \textit{a} to \textit{f} takes place.}
\label{non-linear_term}
\end{figure}

The linear terms are obtained by the variation of densities due to the direct update of nodes and links in the real time steps. As a way of example, now we derive in detail the first term of the first equation: In any update step, with probability $\rho_{a}$ a pair \textit{a} is randomly chosen. According to the update rule Fig.~\ref{update_rule}, with probability $p$ it turns into the pair \textit{b} and changes the global density as $\Delta\rho_a=-\frac{1}{N\mu/2}$. Also, with probability $1-p$, the pair \textit{a} becomes \textit{f} and changes the global density in the same amount. In addition, the time interval (measured in Monte Carlo steps) in any update step is given by $\Delta t=\frac{1}{N\mu/2}$. Therefore, the variation of $\rho_{a}$ due to the direct effect of update of the pair \textit{a} is
\begin{eqnarray}
\frac{d\rho_a}{dt}\Big\rvert_{a\rightarrow b}+\frac{d\rho_a}{dt}\Big\rvert_{a\rightarrow f}&=&p\frac{\Delta\rho_a}{\Delta t}\rho_a+(1-p)\frac{\Delta\rho_a}{\Delta t}\rho_a =-\rho_{a},
\end{eqnarray}
where $\frac{d\rho_i}{dt}\Big\rvert_{i\rightarrow j}$ is the direct effect of the update from the pair $i\in \{a,c,e\}$ to the pair $j\in \{b,d,f\}$. In general
\begin{eqnarray}
\frac{d\rho_i}{dt}=\sum_{j}\frac{d\rho_i}{dt}\Big\rvert_{i\rightarrow j}+\dots, \hspace{20pt}i\in \{a,c,e\}.
\end{eqnarray}
Needless to say that because of conservation of number of pair connection in the network, the negative value of $\frac{d\rho_a}{dt}\Big\rvert_{a\rightarrow b}$ and $\frac{d\rho_a}{dt}\Big\rvert_{a\rightarrow f}$ should be added to the variation of \textit{b} and \textit{f}, respectively. Thus in general
\begin{eqnarray}
\frac{d\rho_{j}}{dt}&=&-\frac{d\rho_i}{dt}\Big\rvert_{i\rightarrow {j}}+\dots\hspace{20pt}j\in \{b,d,f\}.
\end{eqnarray}
However, the non-linear terms are an indirect effect of the node update. The blue nodes can be an end to any of the links \textit{c}, \textit{d}, \textit{e} and \textit{f} while the white nodes can be an end to any of the links \textit{a}, \textit{b}, \textit{e} and \textit{f}. Thus, the node update will change the status of the connected links to the updated node. As case in point, to show how we obtain the nonlinear terms, we derive the first non linear term of the fifth equation $-(1-p)(\mu-1)\frac{\rho_{a}\rho_{e}}{2\chi}$. This term is the implication of a node update from the pair connection \textit{a} to \textit{f}. As mentioned before, in any Monte Carlo step, with probability $\rho_{a}$ a pair \textit{a} is randomly picked and with probability $1-p$ a node update takes place. Now, let us examine the change of the rate of $\rho_{e}$ under the update of \textit{a} to \textit{f} as presented in Fig.~\ref{non-linear_term}. The normalized number of whole pair connections attached to the one side of a pair \textit{a} is $\frac{\mu-1}{N\mu/2}$ and from this portion, the fraction of the pair \textit{e} that is attached to the link \textit{a} is $\frac{\rho_{e}}{\chi}\frac{\mu-1}{N\mu/2}$. In addition, due to the asymmetry on the shape of pairs \textit{e} and \textit{f}, the update from any side of these pairs would result in a different pairs. For instance, if the white opinion in the pair \textit{e} is updated, the new pair convert to \textit{d} and if blue opinion is flipped, the new one turns to \textit{b}. Thus, when we deal with pairs \textit{e} and \textit{f}, the contribution of nonlinear terms in the node update should be considered by probability $\frac{1}{2}$. Thus, the global change in the density of the pair \textit{e} due to the node update from \textit{a} to \textit{f} is given by $\Delta\rho_e=-\frac{\rho_{e}}{2\chi}\frac{\mu-1}{N\mu/2}$.
\begin{eqnarray}
\frac{d\rho_e}{dt}\Big\rvert_{a\to f}^{e\to d}=(1-p)\frac{\Delta\rho_e}{\Delta t}\rho_a =-(1-p)(\mu-1)\frac{\rho_{a}\rho_{e}}{2\chi}
\end{eqnarray}
where $\frac{d\rho_v}{dt}\Big\rvert_{i\to j}^{v\to w}$ and its negative value are the change in density of the pairs $v\in \{a,b,c,d,e,f\}$ and $w\in \{a,b,c,d,e,f\}$, respectively, due to the node update of the pair $i$ to $j$. In general
\begin{eqnarray}
\frac{d\rho_{v}}{dt}&=&...+\frac{d\rho_v}{dt}\Big\rvert_{i\to j}^{v\to w}+... \nonumber \\
\frac{d\rho_{w}}{dt}&=&...-\frac{d\rho_v}{dt}\Big\rvert_{i\to j}^{v\to w}+....
\end{eqnarray}
In this way we are able to obtain all the nonlinear terms of Eqs.(\ref{eq:rate}).

\subsection{Initial condition}
When one integrates numerically Eqs.\ref{eq:rate} it is important to ensure that the initial condition satisfies the relations (\ref{rho_rel}). This is achieved by using as initial condition
\begin{eqnarray}
\rho_a(0)&=&x_0^2(1-\ell_0)\nonumber\\
\rho_b(0)&=&x_0^2\ell_0\nonumber\\
\rho_c(0)&=&(1-x_0)^2(1-\ell_1)\nonumber\\
\rho_d(0)&=&(1-x_0)^2\ell_1 \nonumber\\
\rho_e(0)&=&2x_0(1-x_0)\ell_2 \nonumber\\
\rho_f(0)&=&2x_0(1-x_0)(1-\ell_2)
\end{eqnarray}
where $\ell_0$ and $\ell_1$ are the fraction of friendly links within the groups with white and blue opinion, respectively, and $\ell_2$ is the fraction of friendly links between the white and blue opinion groups. For the sake of simplicity, and otherwise stated, in this work we consider $\ell_0=\ell_1=\ell_2$.

\bibliography{Draft}

\begin{thebibliography}{27}%
\makeatletter
\providecommand \@ifxundefined [1]{%
 \@ifx{#1\undefined}
}%
\providecommand \@ifnum [1]{%
 \ifnum #1\expandafter \@firstoftwo
 \else \expandafter \@secondoftwo
 \fi
}%
\providecommand \@ifx [1]{%
 \ifx #1\expandafter \@firstoftwo
 \else \expandafter \@secondoftwo
 \fi
}%
\providecommand \natexlab [1]{#1}%
\providecommand \enquote  [1]{``#1''}%
\providecommand \bibnamefont  [1]{#1}%
\providecommand \bibfnamefont [1]{#1}%
\providecommand \citenamefont [1]{#1}%
\providecommand \href@noop [0]{\@secondoftwo}%
\providecommand \href [0]{\begingroup \@sanitize@url \@href}%
\providecommand \@href[1]{\@@startlink{#1}\@@href}%
\providecommand \@@href[1]{\endgroup#1\@@endlink}%
\providecommand \@sanitize@url [0]{\catcode `\\12\catcode `\$12\catcode
  `\&12\catcode `\#12\catcode `\^12\catcode `\_12\catcode `\%12\relax}%
\providecommand \@@startlink[1]{}%
\providecommand \@@endlink[0]{}%
\providecommand \url  [0]{\begingroup\@sanitize@url \@url }%
\providecommand \@url [1]{\endgroup\@href {#1}{\urlprefix }}%
\providecommand \urlprefix  [0]{URL }%
\providecommand \Eprint [0]{\href }%
\providecommand \doibase [0]{http://dx.doi.org/}%
\providecommand \selectlanguage [0]{\@gobble}%
\providecommand \bibinfo  [0]{\@secondoftwo}%
\providecommand \bibfield  [0]{\@secondoftwo}%
\providecommand \translation [1]{[#1]}%
\providecommand \BibitemOpen [0]{}%
\providecommand \bibitemStop [0]{}%
\providecommand \bibitemNoStop [0]{.\EOS\space}%
\providecommand \EOS [0]{\spacefactor3000\relax}%
\providecommand \BibitemShut  [1]{\csname bibitem#1\endcsname}%
\let\auto@bib@innerbib\@empty
\bibitem [{\citenamefont {Heider}(1946)}]{heider1946attitudes}%
  \BibitemOpen
  \bibfield  {author} {\bibinfo {author} {\bibfnamefont {F.}~\bibnamefont
  {Heider}},\ }\href@noop {} {\bibfield  {journal} {\bibinfo  {journal} {The
  Journal of psychology}\ }\textbf {\bibinfo {volume} {21}},\ \bibinfo {pages}
  {107} (\bibinfo {year} {1946})}\BibitemShut {NoStop}%
\bibitem [{\citenamefont {Radicchi}\ \emph {et~al.}(2007)\citenamefont
  {Radicchi}, \citenamefont {Vilone}, \citenamefont {Yoon},\ and\ \citenamefont
  {Meyer-Ortmanns}}]{radicchi2007social}%
  \BibitemOpen
  \bibfield  {author} {\bibinfo {author} {\bibfnamefont {F.}~\bibnamefont
  {Radicchi}}, \bibinfo {author} {\bibfnamefont {D.}~\bibnamefont {Vilone}},
  \bibinfo {author} {\bibfnamefont {S.}~\bibnamefont {Yoon}}, \ and\ \bibinfo
  {author} {\bibfnamefont {H.}~\bibnamefont {Meyer-Ortmanns}},\ }\href@noop {}
  {\bibfield  {journal} {\bibinfo  {journal} {Physical Review E}\ }\textbf
  {\bibinfo {volume} {75}},\ \bibinfo {pages} {026106} (\bibinfo {year}
  {2007})}\BibitemShut {NoStop}%
\bibitem [{\citenamefont {Szell}\ \emph {et~al.}(2010)\citenamefont {Szell},
  \citenamefont {Lambiotte},\ and\ \citenamefont
  {Thurner}}]{szell2010multirelational}%
  \BibitemOpen
  \bibfield  {author} {\bibinfo {author} {\bibfnamefont {M.}~\bibnamefont
  {Szell}}, \bibinfo {author} {\bibfnamefont {R.}~\bibnamefont {Lambiotte}}, \
  and\ \bibinfo {author} {\bibfnamefont {S.}~\bibnamefont {Thurner}},\
  }\href@noop {} {\bibfield  {journal} {\bibinfo  {journal} {Proceedings of the
  National Academy of Sciences}\ }\textbf {\bibinfo {volume} {107}},\ \bibinfo
  {pages} {13636} (\bibinfo {year} {2010})}\BibitemShut {NoStop}%
\bibitem [{\citenamefont {Marvel}\ \emph {et~al.}(2011)\citenamefont {Marvel},
  \citenamefont {Kleinberg}, \citenamefont {Kleinberg},\ and\ \citenamefont
  {Strogatz}}]{marvel2011continuous}%
  \BibitemOpen
  \bibfield  {author} {\bibinfo {author} {\bibfnamefont {S.~A.}\ \bibnamefont
  {Marvel}}, \bibinfo {author} {\bibfnamefont {J.}~\bibnamefont {Kleinberg}},
  \bibinfo {author} {\bibfnamefont {R.~D.}\ \bibnamefont {Kleinberg}}, \ and\
  \bibinfo {author} {\bibfnamefont {S.~H.}\ \bibnamefont {Strogatz}},\
  }\href@noop {} {\bibfield  {journal} {\bibinfo  {journal} {Proceedings of the
  National Academy of Sciences}\ }\textbf {\bibinfo {volume} {108}},\ \bibinfo
  {pages} {1771} (\bibinfo {year} {2011})}\BibitemShut {NoStop}%
\bibitem [{\citenamefont {Traag}\ and\ \citenamefont
  {Bruggeman}(2009)}]{traag2009community}%
  \BibitemOpen
  \bibfield  {author} {\bibinfo {author} {\bibfnamefont {V.~A.}\ \bibnamefont
  {Traag}}\ and\ \bibinfo {author} {\bibfnamefont {J.}~\bibnamefont
  {Bruggeman}},\ }\href@noop {} {\bibfield  {journal} {\bibinfo  {journal}
  {Physical Review E}\ }\textbf {\bibinfo {volume} {80}},\ \bibinfo {pages}
  {036115} (\bibinfo {year} {2009})}\BibitemShut {NoStop}%
\bibitem [{\citenamefont {Evans}\ and\ \citenamefont
  {Lambiotte}(2009)}]{evans2009line}%
  \BibitemOpen
  \bibfield  {author} {\bibinfo {author} {\bibfnamefont {T.}~\bibnamefont
  {Evans}}\ and\ \bibinfo {author} {\bibfnamefont {R.}~\bibnamefont
  {Lambiotte}},\ }\href@noop {} {\bibfield  {journal} {\bibinfo  {journal}
  {Physical Review E}\ }\textbf {\bibinfo {volume} {80}},\ \bibinfo {pages}
  {016105} (\bibinfo {year} {2009})}\BibitemShut {NoStop}%
\bibitem [{\citenamefont {Ahn}\ \emph {et~al.}(2010)\citenamefont {Ahn},
  \citenamefont {Bagrow},\ and\ \citenamefont {Lehmann}}]{ahn2010link}%
  \BibitemOpen
  \bibfield  {author} {\bibinfo {author} {\bibfnamefont {Y.-Y.}\ \bibnamefont
  {Ahn}}, \bibinfo {author} {\bibfnamefont {J.~P.}\ \bibnamefont {Bagrow}}, \
  and\ \bibinfo {author} {\bibfnamefont {S.}~\bibnamefont {Lehmann}},\
  }\href@noop {} {\bibfield  {journal} {\bibinfo  {journal} {nature}\ }\textbf
  {\bibinfo {volume} {466}},\ \bibinfo {pages} {761} (\bibinfo {year}
  {2010})}\BibitemShut {NoStop}%
\bibitem [{\citenamefont {Nepusz}\ and\ \citenamefont
  {Vicsek}(2012)}]{nepusz2012controlling}%
  \BibitemOpen
  \bibfield  {author} {\bibinfo {author} {\bibfnamefont {T.}~\bibnamefont
  {Nepusz}}\ and\ \bibinfo {author} {\bibfnamefont {T.}~\bibnamefont
  {Vicsek}},\ }\href@noop {} {\bibfield  {journal} {\bibinfo  {journal} {Nature
  Physics}\ }\textbf {\bibinfo {volume} {8}},\ \bibinfo {pages} {568} (\bibinfo
  {year} {2012})}\BibitemShut {NoStop}%
\bibitem [{\citenamefont {Antal}\ \emph {et~al.}(2005)\citenamefont {Antal},
  \citenamefont {Krapivsky},\ and\ \citenamefont {Redner}}]{antal2005dynamics}%
  \BibitemOpen
  \bibfield  {author} {\bibinfo {author} {\bibfnamefont {T.}~\bibnamefont
  {Antal}}, \bibinfo {author} {\bibfnamefont {P.~L.}\ \bibnamefont
  {Krapivsky}}, \ and\ \bibinfo {author} {\bibfnamefont {S.}~\bibnamefont
  {Redner}},\ }\href@noop {} {\bibfield  {journal} {\bibinfo  {journal}
  {Physical Review E}\ }\textbf {\bibinfo {volume} {72}},\ \bibinfo {pages}
  {036121} (\bibinfo {year} {2005})}\BibitemShut {NoStop}%
\bibitem [{\citenamefont {Marvel}\ \emph {et~al.}(2009)\citenamefont {Marvel},
  \citenamefont {Strogatz},\ and\ \citenamefont
  {Kleinberg}}]{marvel2009energy}%
  \BibitemOpen
  \bibfield  {author} {\bibinfo {author} {\bibfnamefont {S.~A.}\ \bibnamefont
  {Marvel}}, \bibinfo {author} {\bibfnamefont {S.~H.}\ \bibnamefont
  {Strogatz}}, \ and\ \bibinfo {author} {\bibfnamefont {J.~M.}\ \bibnamefont
  {Kleinberg}},\ }\href@noop {} {\bibfield  {journal} {\bibinfo  {journal}
  {Physical review letters}\ }\textbf {\bibinfo {volume} {103}},\ \bibinfo
  {pages} {198701} (\bibinfo {year} {2009})}\BibitemShut {NoStop}%
\bibitem [{\citenamefont {Antal}\ \emph {et~al.}(2006)\citenamefont {Antal},
  \citenamefont {Krapivsky},\ and\ \citenamefont {Redner}}]{antal2006social}%
  \BibitemOpen
  \bibfield  {author} {\bibinfo {author} {\bibfnamefont {T.}~\bibnamefont
  {Antal}}, \bibinfo {author} {\bibfnamefont {P.~L.}\ \bibnamefont
  {Krapivsky}}, \ and\ \bibinfo {author} {\bibfnamefont {S.}~\bibnamefont
  {Redner}},\ }\href@noop {} {\bibfield  {journal} {\bibinfo  {journal}
  {Physica D: Nonlinear Phenomena}\ }\textbf {\bibinfo {volume} {224}},\
  \bibinfo {pages} {130} (\bibinfo {year} {2006})}\BibitemShut {NoStop}%
\bibitem [{\citenamefont {Leskovec}\ \emph {et~al.}(2010)\citenamefont
  {Leskovec}, \citenamefont {Huttenlocher},\ and\ \citenamefont
  {Kleinberg}}]{leskovec2010signed}%
  \BibitemOpen
  \bibfield  {author} {\bibinfo {author} {\bibfnamefont {J.}~\bibnamefont
  {Leskovec}}, \bibinfo {author} {\bibfnamefont {D.}~\bibnamefont
  {Huttenlocher}}, \ and\ \bibinfo {author} {\bibfnamefont {J.}~\bibnamefont
  {Kleinberg}},\ }in\ \href@noop {} {\emph {\bibinfo {booktitle} {Proceedings
  of the SIGCHI conference on human factors in computing systems}}}\ (\bibinfo
  {organization} {ACM},\ \bibinfo {year} {2010})\ pp.\ \bibinfo {pages}
  {1361--1370}\BibitemShut {NoStop}%
\bibitem [{\citenamefont {Fern{\'a}ndez-Gracia}\ \emph
  {et~al.}(2012)\citenamefont {Fern{\'a}ndez-Gracia}, \citenamefont
  {Castell{\'o}}, \citenamefont {Egu{\'\i}luz},\ and\ \citenamefont
  {San~Miguel}}]{fernandez2012dynamics}%
  \BibitemOpen
  \bibfield  {author} {\bibinfo {author} {\bibfnamefont {J.}~\bibnamefont
  {Fern{\'a}ndez-Gracia}}, \bibinfo {author} {\bibfnamefont {X.}~\bibnamefont
  {Castell{\'o}}}, \bibinfo {author} {\bibfnamefont {V.~M.}\ \bibnamefont
  {Egu{\'\i}luz}}, \ and\ \bibinfo {author} {\bibfnamefont {M.}~\bibnamefont
  {San~Miguel}},\ }\href@noop {} {\bibfield  {journal} {\bibinfo  {journal}
  {Physical Review E}\ }\textbf {\bibinfo {volume} {86}},\ \bibinfo {pages}
  {066113} (\bibinfo {year} {2012})}\BibitemShut {NoStop}%
\bibitem [{\citenamefont {Carro}\ \emph {et~al.}(2014)\citenamefont {Carro},
  \citenamefont {Vazquez}, \citenamefont {Toral},\ and\ \citenamefont
  {San~Miguel}}]{carro2014fragmentation}%
  \BibitemOpen
  \bibfield  {author} {\bibinfo {author} {\bibfnamefont {A.}~\bibnamefont
  {Carro}}, \bibinfo {author} {\bibfnamefont {F.}~\bibnamefont {Vazquez}},
  \bibinfo {author} {\bibfnamefont {R.}~\bibnamefont {Toral}}, \ and\ \bibinfo
  {author} {\bibfnamefont {M.}~\bibnamefont {San~Miguel}},\ }\href@noop {}
  {\bibfield  {journal} {\bibinfo  {journal} {Physical Review E}\ }\textbf
  {\bibinfo {volume} {89}},\ \bibinfo {pages} {062802} (\bibinfo {year}
  {2014})}\BibitemShut {NoStop}%
\bibitem [{\citenamefont {Shi}\ \emph {et~al.}(2016)\citenamefont {Shi},
  \citenamefont {Proutiere}, \citenamefont {Johansson}, \citenamefont {Baras},\
  and\ \citenamefont {Johansson}}]{shi2016evolution}%
  \BibitemOpen
  \bibfield  {author} {\bibinfo {author} {\bibfnamefont {G.}~\bibnamefont
  {Shi}}, \bibinfo {author} {\bibfnamefont {A.}~\bibnamefont {Proutiere}},
  \bibinfo {author} {\bibfnamefont {M.}~\bibnamefont {Johansson}}, \bibinfo
  {author} {\bibfnamefont {J.~S.}\ \bibnamefont {Baras}}, \ and\ \bibinfo
  {author} {\bibfnamefont {K.~H.}\ \bibnamefont {Johansson}},\ }\href@noop {}
  {\bibfield  {journal} {\bibinfo  {journal} {Operations Research}\ }\textbf
  {\bibinfo {volume} {64}},\ \bibinfo {pages} {585} (\bibinfo {year}
  {2016})}\BibitemShut {NoStop}%
\bibitem [{\citenamefont {Carro}\ \emph {et~al.}(2016)\citenamefont {Carro},
  \citenamefont {Toral},\ and\ \citenamefont {San~Miguel}}]{carro2016coupled}%
  \BibitemOpen
  \bibfield  {author} {\bibinfo {author} {\bibfnamefont {A.}~\bibnamefont
  {Carro}}, \bibinfo {author} {\bibfnamefont {R.}~\bibnamefont {Toral}}, \ and\
  \bibinfo {author} {\bibfnamefont {M.}~\bibnamefont {San~Miguel}},\
  }\href@noop {} {\bibfield  {journal} {\bibinfo  {journal} {New Journal of
  Physics}\ }\textbf {\bibinfo {volume} {18}},\ \bibinfo {pages} {113056}
  (\bibinfo {year} {2016})}\BibitemShut {NoStop}%
\bibitem [{\citenamefont {Saeedian}\ \emph {et~al.}(2017)\citenamefont
  {Saeedian}, \citenamefont {Azimi-Tafreshi}, \citenamefont {Jafari},\ and\
  \citenamefont {Kertesz}}]{saeedian2017epidemic}%
  \BibitemOpen
  \bibfield  {author} {\bibinfo {author} {\bibfnamefont {M.}~\bibnamefont
  {Saeedian}}, \bibinfo {author} {\bibfnamefont {N.}~\bibnamefont
  {Azimi-Tafreshi}}, \bibinfo {author} {\bibfnamefont {G.}~\bibnamefont
  {Jafari}}, \ and\ \bibinfo {author} {\bibfnamefont {J.}~\bibnamefont
  {Kertesz}},\ }\href@noop {} {\bibfield  {journal} {\bibinfo  {journal}
  {Physical Review E}\ }\textbf {\bibinfo {volume} {95}},\ \bibinfo {pages}
  {022314} (\bibinfo {year} {2017})}\BibitemShut {NoStop}%
\bibitem [{\citenamefont {Singh}\ \emph {et~al.}(2014)\citenamefont {Singh},
  \citenamefont {Dasgupta},\ and\ \citenamefont {Sinha}}]{singh2014extreme}%
  \BibitemOpen
  \bibfield  {author} {\bibinfo {author} {\bibfnamefont {R.}~\bibnamefont
  {Singh}}, \bibinfo {author} {\bibfnamefont {S.}~\bibnamefont {Dasgupta}}, \
  and\ \bibinfo {author} {\bibfnamefont {S.}~\bibnamefont {Sinha}},\
  }\href@noop {} {\bibfield  {journal} {\bibinfo  {journal} {EPL (Europhysics
  Letters)}\ }\textbf {\bibinfo {volume} {105}},\ \bibinfo {pages} {10003}
  (\bibinfo {year} {2014})}\BibitemShut {NoStop}%
\bibitem [{\citenamefont {Holovatch}\ \emph {et~al.}(2017)\citenamefont
  {Holovatch}, \citenamefont {Kenna},\ and\ \citenamefont
  {Thurner}}]{holovatch2017complex}%
  \BibitemOpen
  \bibfield  {author} {\bibinfo {author} {\bibfnamefont {Y.}~\bibnamefont
  {Holovatch}}, \bibinfo {author} {\bibfnamefont {R.}~\bibnamefont {Kenna}}, \
  and\ \bibinfo {author} {\bibfnamefont {S.}~\bibnamefont {Thurner}},\
  }\href@noop {} {\bibfield  {journal} {\bibinfo  {journal} {European Journal
  of Physics}\ }\textbf {\bibinfo {volume} {38}},\ \bibinfo {pages} {023002}
  (\bibinfo {year} {2017})}\BibitemShut {NoStop}%
\bibitem [{\citenamefont {Cartwright}\ and\ \citenamefont
  {Harary}(1956)}]{cartwright1956structural}%
  \BibitemOpen
  \bibfield  {author} {\bibinfo {author} {\bibfnamefont {D.}~\bibnamefont
  {Cartwright}}\ and\ \bibinfo {author} {\bibfnamefont {F.}~\bibnamefont
  {Harary}},\ }\href@noop {} {\bibfield  {journal} {\bibinfo  {journal}
  {Psychological review}\ }\textbf {\bibinfo {volume} {63}},\ \bibinfo {pages}
  {277} (\bibinfo {year} {1956})}\BibitemShut {NoStop}%
\bibitem [{\citenamefont {Kermack}\ and\ \citenamefont
  {McKendrick}(1932)}]{kermack1932contributions}%
  \BibitemOpen
  \bibfield  {author} {\bibinfo {author} {\bibfnamefont {W.~O.}\ \bibnamefont
  {Kermack}}\ and\ \bibinfo {author} {\bibfnamefont {A.~G.}\ \bibnamefont
  {McKendrick}},\ }\href@noop {} {\bibfield  {journal} {\bibinfo  {journal}
  {Proc. R. Soc. Lond. A}\ }\textbf {\bibinfo {volume} {138}},\ \bibinfo
  {pages} {55} (\bibinfo {year} {1932})}\BibitemShut {NoStop}%
\bibitem [{\citenamefont {Kermack}\ and\ \citenamefont
  {McKendrick}(1933)}]{kermack1933contributions}%
  \BibitemOpen
  \bibfield  {author} {\bibinfo {author} {\bibfnamefont {W.~O.}\ \bibnamefont
  {Kermack}}\ and\ \bibinfo {author} {\bibfnamefont {A.~G.}\ \bibnamefont
  {McKendrick}},\ }\href@noop {} {\bibfield  {journal} {\bibinfo  {journal}
  {Proc. R. Soc. Lond. A}\ }\textbf {\bibinfo {volume} {141}},\ \bibinfo
  {pages} {94} (\bibinfo {year} {1933})}\BibitemShut {NoStop}%
\bibitem [{\citenamefont {Barrat}\ \emph {et~al.}(2008)\citenamefont {Barrat},
  \citenamefont {Barthelemy},\ and\ \citenamefont
  {Vespignani}}]{barrat2008dynamical}%
  \BibitemOpen
  \bibfield  {author} {\bibinfo {author} {\bibfnamefont {A.}~\bibnamefont
  {Barrat}}, \bibinfo {author} {\bibfnamefont {M.}~\bibnamefont {Barthelemy}},
  \ and\ \bibinfo {author} {\bibfnamefont {A.}~\bibnamefont {Vespignani}},\
  }\href@noop {} {\emph {\bibinfo {title} {Dynamical processes on complex
  networks}}}\ (\bibinfo  {publisher} {Cambridge university press},\ \bibinfo
  {year} {2008})\BibitemShut {NoStop}%
\bibitem [{\citenamefont {Strogatz}(2018)}]{strogatz2018nonlinear}%
  \BibitemOpen
  \bibfield  {author} {\bibinfo {author} {\bibfnamefont {S.~H.}\ \bibnamefont
  {Strogatz}},\ }\href@noop {} {\emph {\bibinfo {title} {Nonlinear dynamics and
  chaos: with applications to physics, biology, chemistry, and engineering}}}\
  (\bibinfo  {publisher} {CRC Press},\ \bibinfo {year} {2018})\BibitemShut
  {NoStop}%
\bibitem [{\citenamefont {Hinrichsen}(2000)}]{hinrichsen2000non}%
  \BibitemOpen
  \bibfield  {author} {\bibinfo {author} {\bibfnamefont {H.}~\bibnamefont
  {Hinrichsen}},\ }\href@noop {} {\bibfield  {journal} {\bibinfo  {journal}
  {Advances in physics}\ }\textbf {\bibinfo {volume} {49}},\ \bibinfo {pages}
  {815} (\bibinfo {year} {2000})}\BibitemShut {NoStop}%
\bibitem [{\citenamefont
  {Axelrod}(1997{\natexlab{a}})}]{axelrod1997dissemination}%
  \BibitemOpen
  \bibfield  {author} {\bibinfo {author} {\bibfnamefont {R.}~\bibnamefont
  {Axelrod}},\ }\href@noop {} {\bibfield  {journal} {\bibinfo  {journal}
  {Journal of conflict resolution}\ }\textbf {\bibinfo {volume} {41}},\
  \bibinfo {pages} {203} (\bibinfo {year} {1997}{\natexlab{a}})}\BibitemShut
  {NoStop}%
\bibitem [{\citenamefont {Axelrod}(1997{\natexlab{b}})}]{Axelrod2}%
  \BibitemOpen
  \bibfield  {author} {\bibinfo {author} {\bibfnamefont {R.}~\bibnamefont
  {Axelrod}},\ }\href@noop {} {\emph {\bibinfo {title} {The Complexity of
  Cooperation}}}\ (\bibinfo  {publisher} {Princeton University Press, Princeton
  NJ.},\ \bibinfo {year} {1997})\BibitemShut {NoStop}%
\end{thebibliography}%

\section*{Acknowledgements}
We acknowledge financial support from Agencia Estatal de Investigaci\'on (AEI, Spain) and Fondo Europeo de Desarrollo Regional under Project ESoTECoS Grant No. FIS2015-63628-C2-2-R (AEI/FEDER,UE) and the Spanish State Research Agency, through the Mar{\'\i}a de Maeztu Program for Units of Excellence in R\&D (MDM-2017-0711).

%

\end{document}